\documentclass[aps,pra,superscriptaddress,twocolumn]{revtex4}

\usepackage{comment}
\usepackage[utf8]{inputenc}
\usepackage{amssymb}
\usepackage{amsmath}
\usepackage{amsfonts}
\usepackage{bm}
\usepackage{mathtools}
\usepackage{bbold}
\usepackage{upgreek}
\usepackage{xfrac}
\usepackage{soul}
\usepackage{bm}
\usepackage{comment}
\usepackage{xcolor}
\usepackage{mathtools}

\usepackage[colorlinks=true,citecolor=blue]{hyperref}
\hypersetup{colorlinks=true,citecolor=blue,linkcolor=blue,urlcolor=blue}
\usepackage{url}

\usepackage{txfonts}
\DeclareSymbolFont{matha}{OML}{txmi}{m}{it}
\DeclareMathSymbol{\varv}{\mathord}{matha}{118}
\usepackage{bbm}
\begin{document}

\title{Quantum Quantile Mechanics: Solving Stochastic Differential Equations for Generating Time-Series}

\author{Annie E. Paine}
\affiliation{Qu \& Co B.V., PO Box 75872, 1070 AW, Amsterdam, The Netherlands}
\affiliation{Department of Physics and Astronomy, University of Exeter, Stocker Road, Exeter EX4 4QL, United Kingdom}

\author{Vincent E. Elfving}
\affiliation{Qu \& Co B.V., PO Box 75872, 1070 AW, Amsterdam, The Netherlands}

\author{Oleksandr Kyriienko}
\affiliation{Qu \& Co B.V., PO Box 75872, 1070 AW, Amsterdam, The Netherlands}
\affiliation{Department of Physics and Astronomy, University of Exeter, Stocker Road, Exeter EX4 4QL, United Kingdom}

\date{\today}

\begin{abstract}
We propose a quantum algorithm for sampling from a solution of stochastic differential equations (SDEs). Using differentiable quantum circuits (DQCs) with a feature map encoding of latent variables, we represent the quantile function for an underlying probability distribution and extract samples as DQC expectation values. Using quantile mechanics we propagate the system in time, thereby allowing for time-series generation. We test the method by simulating the Ornstein-Uhlenbeck process and sampling at times different from the initial point, as required in financial analysis and dataset augmentation. Additionally, we analyse continuous quantum generative adversarial networks (qGANs), and show that they represent quantile functions with a modified (reordered) shape that impedes their efficient time-propagation. Our results shed light on the connection between quantum quantile mechanics (QQM) and qGANs for SDE-based distributions, and point the importance of differential constraints for model training, analogously with the recent success of physics informed neural networks.
\end{abstract}

\maketitle

\begin{flushleft}\textbf{INTRODUCTION}\end{flushleft}\vspace{-2mm}
Stochastic differential equations (SDEs) describe a broad range of phenomena. They emerge when dealing with Brownian motion and quantum noise~\cite{Gardiner2000}. In physical sciences, SDEs are used for describing quantum dynamics~\cite{Breuer2002}, thermal effects \cite{VanKampen1976}, molecular dynamics \cite{Leimkuhler2015}, and they lie at the core of stochastic fluid dynamics~\cite{Holm2015,Cintolesi2020}. In biology SDEs help in the studies of population dynamics \cite{Dobramysl2018} and epidemiology \cite{Allen2008,Allen2017}. They can also help to detect anomalies~\cite{Rajabzadeh2016}. SDEs are widely used in financial calculus~\cite{OksendalBook}, a fundamental component of all mechanisms of pricing financial derivatives and description of market dynamics. Potential applications of SDEs in finance lie in predicting stock prices, currency exchange rates and more ~\cite{Leung2015}. 

A general system of stochastic differential equations 
can be written as~\cite{OksendalBook}
\begin{align}
\label{eq:SDE}
    d \bm{X}_t = f(\bm{X}_t, t) dt + g(\bm{X}_t, t) dW_t,
\end{align}
where $\bm{X}_t$ is a vector of stochastic variables parameterized by time $t$ (or other parameters). Deterministic functions $f$ and $g$ correspond to the drift and diffusion processes, respectively. $W_t$ corresponds to the stochastic Wiener process. 
The stochastic component makes SDEs distinct from other types of partial differential equations, adding a non-differentiable contribution. This also makes SDEs generally difficult to treat. 
Several questions arise: how do we solve Eq.~\eqref{eq:SDE}, and what kind of information do we want to get by solving SDEs? These are not trivial questions to answer, because one might be interested in different aspects of SDE-modelled processes. 

First, the system of SDEs can be rewritten in the form of a partial differential equation for the underlying probability density function (PDF) $p(\bm{x},t)$ of deterministic continuous variables, and then solved for the specified boundary conditions. The resulting equation is known as a Fokker-Planck (FP) or Kolmogorov equation \cite{Bogachev2015}. It can be exploited for calculating observables (averages) from $p(\bm{x},t)$; in many cases these are the mean and the variance for the stochastic variables, $\mathrm{E}[\bm{X}_t] = \int \bm{x} p(\bm{x},t)d\bm{x}$ and $\mathrm{Var}[\bm{X}_t] = \mathrm{E}[\bm{X}_t^2] - \mathrm{E}[\bm{X}_t]^2$, respectively. Here, a challenge arises when a well-defined boundary condition is missing, and instead some additional data is available. The task then falls into the category of data-driven machine learning problems, which has attracted attention recently \cite{Chen2019,Li2020}.
Also, a (potentially implicit) solution of the FP equation does not offer strategies to generate samples directly. Namely, drawing $\bm{x} \sim p(\bm{x},t)$ from a complicated multidimensional distribution, at different $t$, represents another computationally expensive problem to solve.

Second, Eq.~\eqref{eq:SDE} can be integrated using the Euler-Maruyama method \cite{Kloeden1992}, where the deterministic part of the differential equation is solved with stepwise propagation, and the Wiener process is modelled with a random number generator. This corresponds to the ensemble generation process (broadly speaking, \emph{generative modelling}), which is computationally challenging for complex models. For multidimensional systems this is hindered by the curse of dimensionality. The task of solving random partial differential equations was recently addressed with deep learning using physics-informed neural networks~\cite{zhang2019learning, zhang2019108850, nabian2019}. 
Unsupervised generative modelling often requires using adversarial training \cite{yang2018physicsinformed,Kidger2021} and generative adversarial network (GAN) architecture \cite{Goodfellow2014}.
Moreover, in cases where the information about system parameters is limited, or exemplary datasets are generated from measurement and not known initial conditions, we arrive at the task of equation discovery~\cite{Brunton2016,Schaeffer2017,Raissi2018,Maslyaev2019,Reinbold2019}. 

Quantum computers operate in a multidimensional state space and are intrinsically probabilistic. They offer computational advantage for sampling tasks \cite{Aaronson2013,Lund2017,Arrazola2017,Deshpande2021}. This was demonstrated experimentally with superconducting~\cite{Arute2019} and optical~\cite{JWPan2020} quantum devices. Thus, quantum computing may be a prime candidate for disrupting the field of generative modelling. Yet, the challenge of performing the quantum generative modelling for practical tasks remains open. Ideally, discrete distributions can be loaded into a quantum register using an amplitude encoding $\sum_{k=1}^{K} u_k |k\rangle$, where amplitudes $u_k$ contain information about the probability distribution and $|k\rangle$ are binary states for $\log(K)$-qubit register. This allows for a quadratic speed-up when using the amplitude amplification protocol~\cite{Goldman2021}, and has numerous financial use-cases~\cite{Egger2020}. However, representing arbitrary distributions in an amplitude-encoded form may be difficult (no scalable scheme exists yet), and the processing requires a large-scale fault-tolerant quantum computer \cite{Biamonte2017}. 
A strategy that is viable for near- and mid-term QC relies on variational quantum algorithms based on parametrized quantum circuits \cite{Benedetti2019rev,Cerezo2020rev}. These can be seen as quantum neural networks (QNNs)~\cite{Schuld2019a,Mari2020,Abbas2021,Sam2021,Sam2020}. To date, several variational protocols for solving differential equations have been proposed for ordinary and partial differential equations~\cite{Lubasch2020,Kyriienko2021,Knudsen2020,Garcia-Molina2021}, targeting near-term devices. For stochastic differential equations several algorithms have been explored that are based on wavefunction-based probability encoding ~\cite{Rebentrost2018,Stamatopoulos2020,Kubo2021,Gonzalez-Conde2021,KumarRadha2021} and rely on complicated circuits. A strategy for near- and mid-term devices, SDE-based generative modelling can be addressed by data-based learning of probability distributions.

Motivated by classical GAN successes \cite{Goodfellow2014,Gui2020}, various protocols for the adversarial training of generative quantum models were proposed, and coined as \emph{qGANs}~\cite{Lloyd2018,Dallaire2018}. Here, several sample generation strategies can be employed. 
One possible strategy relies on the variational wavefunction preparation as a distribution proxy, with the sample readout following the Born rule. This corresponds to a quantum circuit Born machine (QCBM) generator~\cite{JGLiu2018}. In this case qGAN was applied to sampling from discrete distributions~\cite{Zeng2019,Benedetti2019b,Du2017,Zoufal2019,Situ2020,Chang2021,Niu2021}, preparing arbitrary quantum states~\cite{Benedetti2019a,Braccia2021}, and being demonstrated experimentally with superconducting circuits for image generation~\cite{Huang2020}. In this setting the sampling procedure is efficient, but its power depends on the register width and the generator is difficult to train at increasing scale. The latter comes from demanding requirements on the training set~\cite{Romero2021} and barren plateaus for global cost functions~\cite{Cerezo2021}. We also note that QCBM can be trained in other ways, including maximum mean discrepancy and various statistical divergences \cite{JGLiu2018,Coyle2020,Coyle2021,Kondratyev2020}, with the goal of preparing distributions arising in financial applications.
The second possible strategy relies on a QNN-based generator representing a \emph{continuous qGAN}~\cite{Romero2021}, recently demonstrated experimentally~\cite{Anand2021}, where a feature map embedding \cite{Goto2020} is used for continuous latent variable distribution. In this case the sampling rate is reduced due to the expectation value measurement, but the model becomes trainable, and its power is ultimately limited by the properties of feature maps and the adversarial training schedule (i.e. minimax game instead of loss minimization). The operation of qGANs was recently surveyed in Ref.~\cite{TongLi2020}.
Finally, other generator architectures are represented by quantum Boltzmann machines~\cite{Amin2018,Du2020,Liu2020} where sampling is based on thermal states of quantum Ising Hamiltonian, or recurrent neural networks for NMR quantum computers~\cite{Takaki2021}.

We note that the prior art in both classical and quantum generative modelling is either concentrated on representing probability density functions in model-driven approaches, or training the generator in a black-box fashion [e.g. (q)GANs] where no physical constraints are added. In this work, we suggest to shift the focus to the very tool that enables sampling --- the \emph{quantile function} (QF) (not to be mistaken with the quantile of a distribution). We propose to solve SDEs by rewriting them as differential equations for the quantile function and using their neural representation (classical or quantum neural networks). In the following, we use feature maps and differentiable quantum circuits (DQCs) to represent directly the quantile function of the probability distribution of the underlying SDE, and propagate them in time by solving the differential equations of quantile mechanics. This allows us to prepare a QNN-based generator that is trained from available data and yet is model-informed. While the expectation-based readout associated to the QNN structure requires multiple shots, the proposed approach has improved trainability compared to QCBM architecture and can work with sparse training data. Specifically, we benchmark the developed quantum quantile mechanics (QQM) approach \cite{qqm_approach} using the Ornstein-Uhlenbeck model (a prototypical stochastic process being the base of many financial calculations). We show how to train QF from data and/or the known model at the initial point of time, and find a time-propagated QF that enables high-quality sampling. We then proceed to show that adversarial schemes (continuous GAN or qGAN) in fact train generators as a \emph{reordered quantile function}. Analyzing the similarities and differences between the two methods, we find that quantile functions in their original meaning are suitable for time propagation, while reordered QFs have apparent difficulties for the task (see Discussion). Our work uncovers the possibilities for time series generation and \emph{data-augmentation} enhanced by quantum resources.
\begin{figure}
\begin{center}
\includegraphics[width=1.0\linewidth]{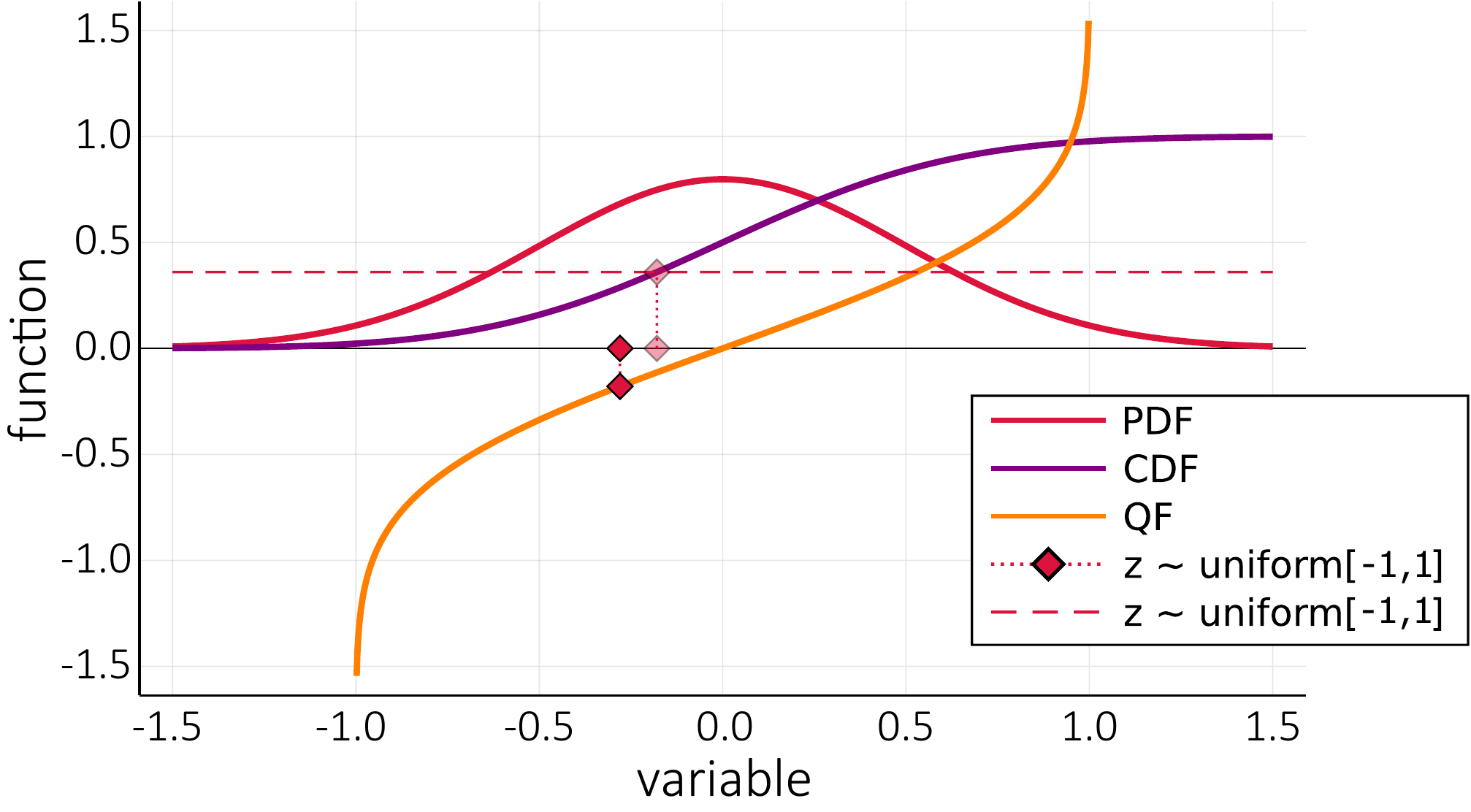}
\end{center}
\caption{\textbf{Illustrative example of sampling.} We plot a normal probability density function as a red curve (PDF), being the Gaussian function with $(\mu, \sigma) = (0, 1/2)$. The domain is chosen as $\mathcal{X} = [-1.5, 1.5]$. The corresponding cumulative distribution function is presented by the purple solid curve (CDF). The quantile function for the corresponding CDF is shown in orange (QF). The red diamonds correspond to a randomly drawn latent variable $z$, and associated probability for the sample value, connected by dotted lines.}
\label{fig:quantile}
\end{figure}



\begin{flushleft}\textbf{RESULTS}\end{flushleft}\vspace{-2mm}
We start by describing the background for generative modelling from SDEs, and proceed to introduce the proposed QQM method. In the second part of the Results section we present numerical experiments for training DQCs to represent time-dependent quantile functions, as well as qGAN training.\vspace{1mm}

\noindent\textbf{Classical sampling and quantile mechanics}

\noindent Let us recall how a sample from a distribution can be obtained using a basic inversion sampling. In the following we consider a single stochastic process $X_t$ (also written as $X$ for brevity), and the generalization to multiple processes/dimensions is straightforward.
First, we take a PDF $p(x)$ of a continuous variable $x \in \mathcal{X}$ being normalized over domain $\mathcal{X}$. Next, we find a cumulative distribution function (CDF) for the stochastic variable $X$ defined as an integral $F_X (x) = \int_{-\infty}^{x} p(x') dx'$. This maps $x$ to the probability value lying in $[0,1]$ range on the ordinate axis (see Fig.~\ref{fig:quantile} for an illustration). 
Being a non-decreasing function, $F_X(x)$ has to be inverted to provide a sample. Generating a random number $z \in (-1,1)$ from the uniform distribution, we can solve the equation $z = F_X (x)$ and find the corresponding sample. This requires finding the inverse of CDF, $F_X^{-1}(x)$. In most cases this leads to a transcendental problem without a closed-form solution. This poses computational challenges and requires graphical solution methods \cite{BoydBook}. Finally, each random number $z \sim \mathtt{uniform(-1,1)}$ gives a random sample $X$ from PDF of interest as $X = F_X^{-1}(z)$ (note that the range of $z$ can be easily rescaled). The inverted CDF function is known as a \emph{quantile function} of the continuous distribution, $F_X^{-1}(z) \equiv Q(z)$.

While generally quantile functions are difficult to get from PDFs, they can be obtained by solving nonlinear partial differential equations derived from SDEs of interest. This approach is called the \emph{quantile mechanics}. The quantile function $Q(z,t)$ can be obtained for any general SDE in the form \eqref{eq:SDE} and its evolution is given by quantilized FP as (see full derivation in Ref.~\cite{Shaw2008})
\begin{align}
\label{eq:QE}
    \frac{\partial Q(z,t)}{\partial t} = f(Q, t) - \frac{1}{2} \frac{\partial g^2(Q, t)}{\partial Q} + \frac{g^2(Q,t)}{2} \left( \frac{\partial Q}{\partial z} \right)^{-2} \frac{\partial^2 Q}{\partial z^2},
\end{align}
where $f(Q,t)$ and $g(Q,t)$ are the drift and diffusion terms familiar from Eq.~\eqref{eq:SDE}. Eq.~\eqref{eq:QE} is solved as a function of latent variable $z$ and time $t$. Once $Q(z,t)$ is known, evaluating it at random uniform $z$'s as $t$ progresses we can get full time series (trajectories) obeying the stochastic differential equation \eqref{eq:SDE}. In Supplemental Information we also consider the case of solving reverse-time SDEs.

\begin{figure}[t]
\begin{center}
\includegraphics[width=1.0\linewidth]{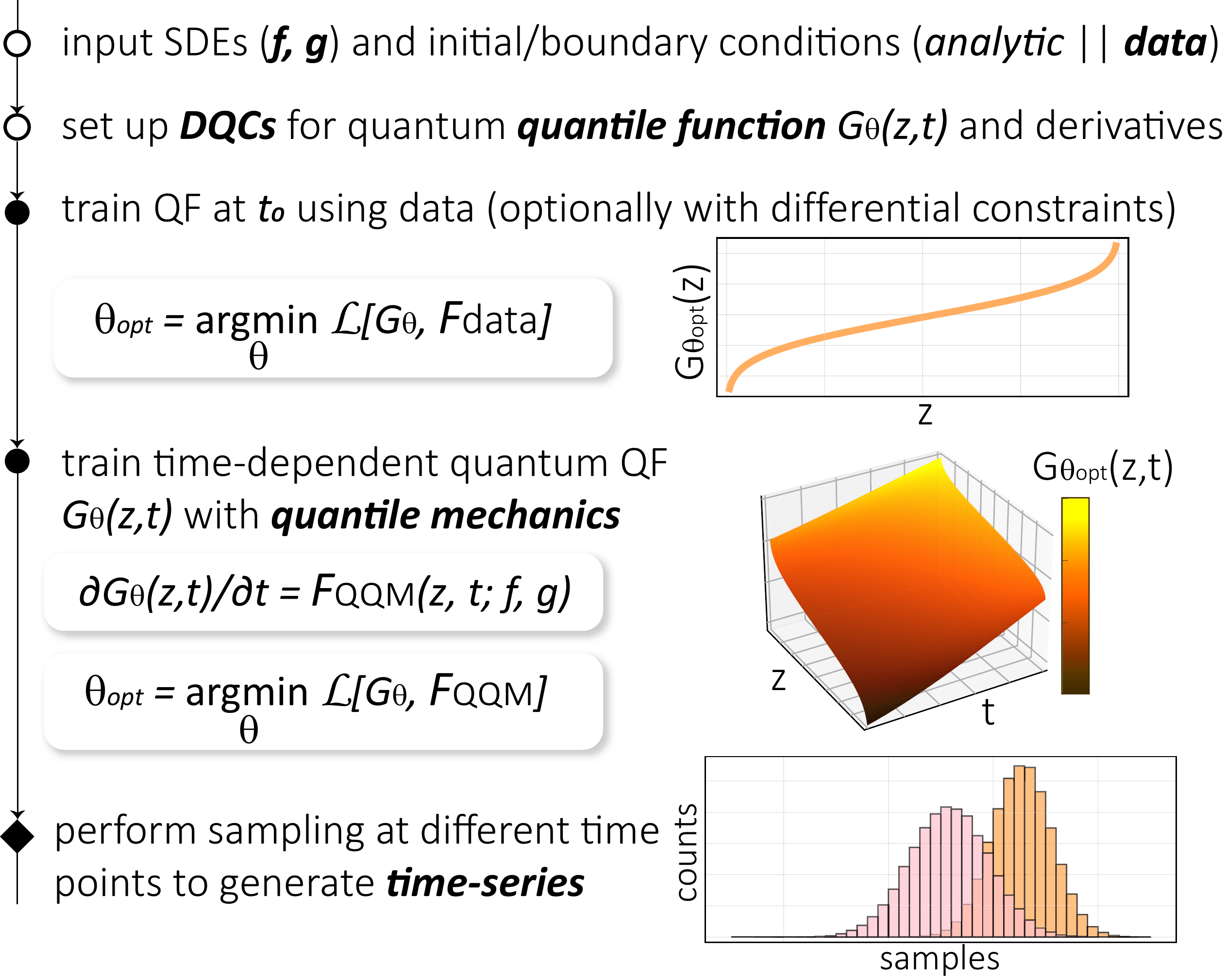}
\end{center}
\caption{\textbf{QQM workflow.} Given the system of stochastic differential equations and initial data, differentiable quantum circuits are trained to represent the corresponding quantile function $G_{\theta}(z,t)$ as a function of a random latent variable ($z$) and propagate it in time ($t$) with quantum mechanics equations. The hybrid quantum-classical loop is used for optimizing variational parameters $\theta$ through loss function $\mathcal{L}$ minimization based on data and differential equations for the initial and propagated QF. Evaluating $G_{\theta_{\mathrm{opt}}}(z \sim \mathtt{uniform(-1,1)},t)$ at optimal angles, random values of the latent variable, and different time points $t$, we generate time series from SDE.}
\label{fig:workflow}
\end{figure}

In general, quantile mechanics equations are not easy to solve. Power series solution as function approximation are known \cite{Shaw2008,CarilloToscani,Shaw2006,Shaw2014}, as well as some simple examples \cite{GilchristBook}. However, the difficulty arises when multidimensional problems are considered and generative modelling suffers from the curse of dimensionality. To harness the full power of quantile mechanics, we thus propose to use neural representation of QFs. To our knowledge, this is the first application of machine learning methods to use quantile-based sampling, and we envisage that both classical and quantum ML can be used for the universal function approximation \cite{universalfunction, Goto2020}. Here, the use of quantum neural networks offers a potential to reproducing complex functions in the high-dimensional space, including systems where strong correlations are important.
In the following, we combine quantum computing and quantile mechanics to develop the \emph{quantum quantile mechanics} approach. The sketch of the QQM workflow is shown in Fig.~\ref{fig:workflow}. We represent QFs as quantum neural networks in the DQC form, thus exploiting the large expressivity of quantum-based learning. 
We show how differential equations for quantile functions can be used for training differentiable quantum circuits. Second, we introduce the quantum quantile learning protocol for inferring QF from data and use QQM to propagate the system in time. This provides a robust protocol for time series generation and sampling. Finally, we show that generative adversarial networks act as quantile functions for randomized association of the latent variable values and samples.\vspace{1mm} 


\noindent\textbf{Quantum quantile mechanics}

\begin{figure}
\begin{center}
\includegraphics[width=1.0\linewidth]{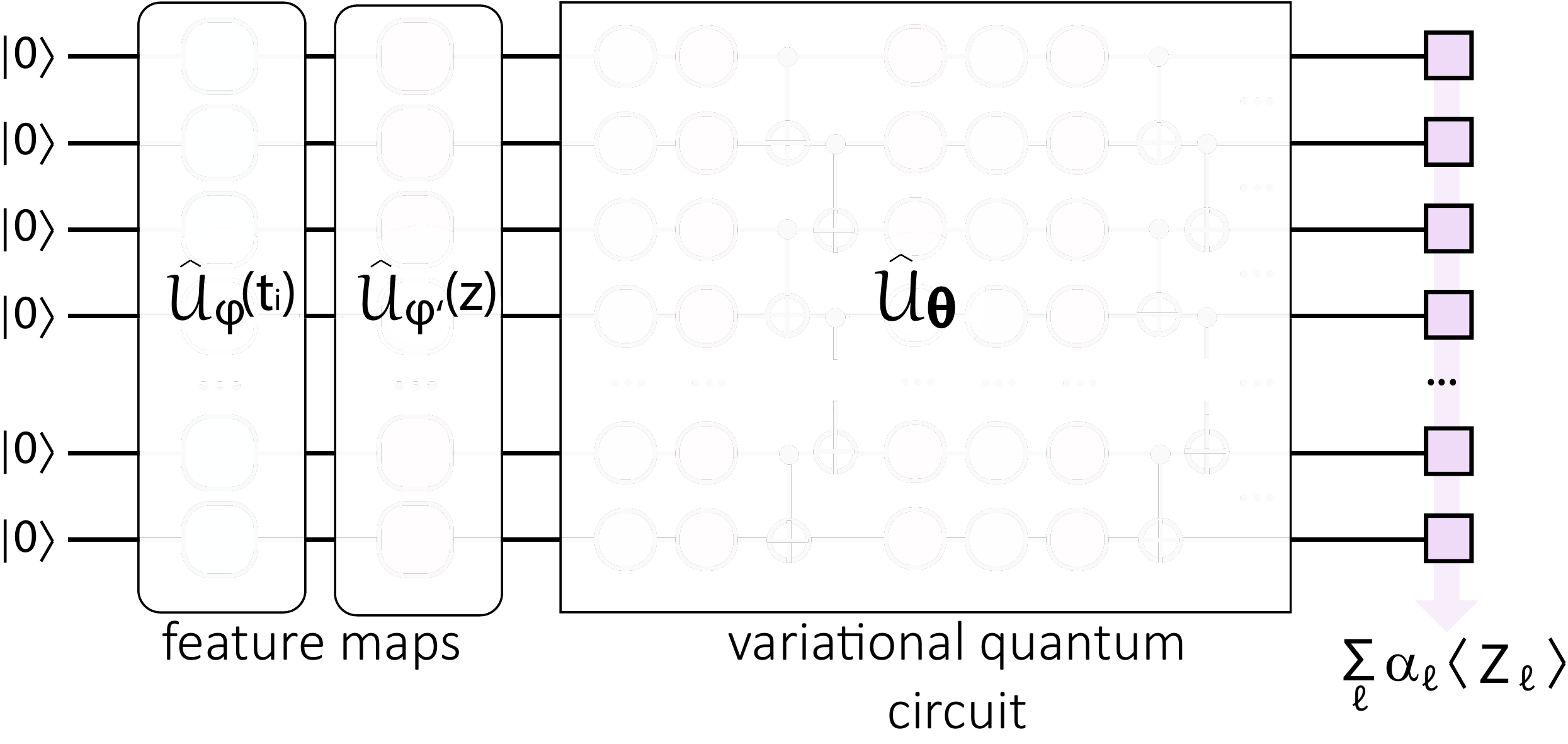}
\end{center}
\caption{\textbf{DQC layout.} State preparation circuit for representing a time-dependent quantile function. Acting on the initial state, two feature maps, with $\hat{R}_y$ using for time-dependence embedding and $\hat{R}_x$ layer for the latent variable embedding. We use HEA for the variational search, and total Z magnetization as a cost function.}
\label{fig:ansatz}
\end{figure}

\noindent To represent a trainable (neural) quantile function we construct a parametrized quantum circuit using quantum embedding through feature maps $\hat{\mathcal{U}}_{\phi}(x)$ (with $\phi$ labelling a mapping function) \cite{Mitarai2018,Schuld2019a,Benedetti2019rev}, followed by variationally-adjustable circuit (ansatz) $\hat{\mathcal{U}}_{\bm{\theta}}$ parametrized by angles $\bm{\theta}$. The latter is routinely used in variational quantum algorithms~\cite{Cerezo2020rev}, and for ML problems has the hardware efficient ansatz (HEA) structure~\cite{Kandala2017}. The power of quantum feature maps comes from mapping $x \in \mathcal{X}$ from the data space to the quantum state $|\psi(x)\rangle = \hat{\mathcal{U}}_{\phi}(x)|{\O}\rangle$ living in the Hilbert space. Importantly, the automatic differentiation of quantum feature maps allows derivatives to be represented as DQCs~\cite{Kyriienko2021}. The readout is set as a sum of weighted expectation values~\cite{Goto2020}. Following this strategy, we assign a generator circuit $G(z,t)$ to represent a function parametrized by $t$ (labels time as before), and the embedded latent variable $z$. 
The generator reads 
\begin{align}
\label{eq:Gzt}
    G(z, t) = \langle {\O} | \hat{\mathcal{U}}_{\phi}(t)^\dagger \hat{\mathcal{U}}_{\phi '}(z)^\dagger \hat{\mathcal{U}}_{\bm{\theta}}^\dagger \Big( \sum_{\ell=1}^{L} \alpha_\ell \hat{\mathcal{C}}_\ell \Big) \hat{\mathcal{U}}_{\bm{\theta}} \hat{\mathcal{U}}_{\phi '}(z) \hat{\mathcal{U}}_{\phi}(t) | {\O} \rangle,
\end{align}
where $\hat{\mathcal{U}}_{\phi}(z)$ and $\hat{\mathcal{U}}_{\phi '}(t)$ are quantum feature maps (possibly different), $\{ \hat{\mathcal{C}}_\ell \}_{\ell=1}^{L}$ represent $L$ distinct Hermitian cost function operators, and $\{ \alpha_\ell \}_{\ell=1}^{L}$ and $\bm{\theta}$ are real coefficients that may be adjusted variationally. We denote the initial state as $| \text{\O} \rangle$, typically chosen as a product state $|0\rangle^{\otimes N}$. The circuit structure is shown in Fig.~\ref{fig:ansatz}. We can also work with multiple latent variables and thus multidimensional distributions.

Our next step is developing the training procedure for $G$ [or specifically for the generator's operator $\hat{\mathcal{U}}_G(z) = \hat{\mathcal{U}}_{\bm{\theta}} \hat{\mathcal{U}}_{\phi '}(z)$], such that it represents QF for an underlying data distribution. Namely, we require that the circuit maps the latent variable $z \in [-1,1]$ 
to a sample $G(z) = Q(z)$.
Choosing different sets of cost functions, with the same quantum circuits, we can produce samples from multidimensional PDF.

The training requires a dataset $\{ X_{\mathrm{data}} \}$ generated from a probability distribution for the system we want to study (or measured experimentally), which serves as an initial/boundary condition. Additionally, we know the underlying processes that describe the system and which serve as differential constraints. The problem is specified by SDEs $d \bm{X}_t = f_{\bm{\xi}}(\bm{X}_t, t) dt + g_{\bm{\xi}}(\bm{X}_t, t) dW_t$, where we explicitly state that the drift and diffusion functions $f_{\bm{\xi}}(\bm{X}_t, t)$ and $g_{\bm{\xi}}(\bm{X}_t, t)$ are parametrized by the vector $\bm{\xi}$ (time-independent). We consider that the SDE parameters for generating data similar to $\{ X_{\mathrm{data}} \}$ are known in the first approximation $\bm{\xi}^{(0)}$. This can be adjusted during the training to have the best convergence, as used in the equation discovery approach~\cite{Brunton2016,Raissi2018}.
The loss is constructed as a sum of data-based and SDE-based contributions, $\mathcal{L} = \mathcal{L}_{\mathrm{data}} + \mathcal{L}_{\mathrm{SDE}}$. The first part $\mathcal{L}_{\mathrm{data}}$ is designed such that the data points in $\{ X_{\mathrm{data}} \}$ are represented by the trained QF. For this, the data is binned appropriately and collected in ascending order, as expected for any quantile function. Then, we use quantum circuit learning (QCL) as a quantum nonlinear regression method~\cite{Mitarai2018} to learn the quantile function. We note that such approach represents a data-frugal strategy, where the need for training on \emph{all} data points is alleviated.
The second loss term $\mathcal{L}_{\mathrm{SDE}}$ is designed such that the learnt quantile function obeys probability models associated to SDEs. Specifically, the generator $G(z,t)$ needs to satisfy the quantilized Fokker-Planck equation~\eqref{eq:QE}. The differential loss is introduced using the DQC approach~\cite{Kyriienko2021}, and reads 
\begin{align}
\label{eq:sde_loss}
    \mathcal{L}_{\mathrm{SDE}} = \frac{1}{M} \sum_{z,t \in \mathcal{Z}, \mathcal{T}} \mathfrak{D}\left[ \frac{\partial G_{\bm{\theta}}}{\partial t}, F(z, t, f, g, \frac{\partial G_{\bm{\theta}}}{\partial z}, \frac{\partial^2 G_{\bm{\theta}}}{\partial z^2}) \right],
\end{align}
where $\mathfrak{D}[a, b]$ denotes the distance measure for two scalars, and the loss is estimated over the grid of points in sets $(\mathcal{Z}, \mathcal{T})$. Here $M = \mathrm{car}(\mathcal{Z}) \mathrm{car}(\mathcal{T})$ is the total number of points. We also introduce the function $F(z, t, f, g, \partial_z G_{\bm{\theta}}, \partial_{zz}^2 G_{\bm{\theta}})$ denoting the RHS for Eq.~\eqref{eq:QE}, or any other differential constraint. 
Other important ingredients of the QQM method include the calculation of second-order derivatives for the feature map encoded functions (as required is $\mathcal{L}_{\mathrm{SDE}}$), and the proposed treatment of initial/boundary conditions for multivariate function. We describe these technical details in the Methods section. Once the training is set up, the loss is minimized using a hybrid quantum-classical loop where optimal variational parameters $\bm{\theta}_{\mathrm{opt}}$ (and $\bm{\alpha}_{\mathrm{opt}}$) are searched using non-convex optimization methods.\vspace{1mm}


\noindent\textbf{QQM-based generative modelling}

\noindent In the next subsections we present numerical simulations of generative modelling. In the first part we apply the developed quantum quantile mechanics approach for solving a specific SDE, and demonstrate a data-enabled operation. In the second part we cover the so-called quantum generative adversarial network (qGAN) that was previously used for continuous distributions, and show numerical results for solving the same problem. The two approaches are then compared in the next section (Discussion).\vspace{1mm}

\textit{Ornstein-Uhlenbeck for financial forecasting and trading.} 
To validate the QQM approach and perform time series forecasting, we pick a prototypical test problem. As an example we choose the Ornstein-Uhlenbeck (OU) process. In financial analysis its generalization is known as the Vasicek model~\cite{Vasicek1977}. It describes the evolution of interest rates and bond prices~\cite{Mamon2004}. This stochastic investment model is the time-independent drift version of the Hull–White model~\cite{HullWhite1990} widely used for derivatives pricing. We also note that OU describes dynamics of currency exchange rates, and is commonly used in Forex pair trading --- a primary example for quantum generative modelling explored to date~\cite{Coyle2021,Kondratyev2020}. Thus, by benchmarking the generative power of QQM for OU we can compare it to other strategies (valid at fixed time point).
\begin{figure*}
\begin{center}
\includegraphics[width=0.95\linewidth]{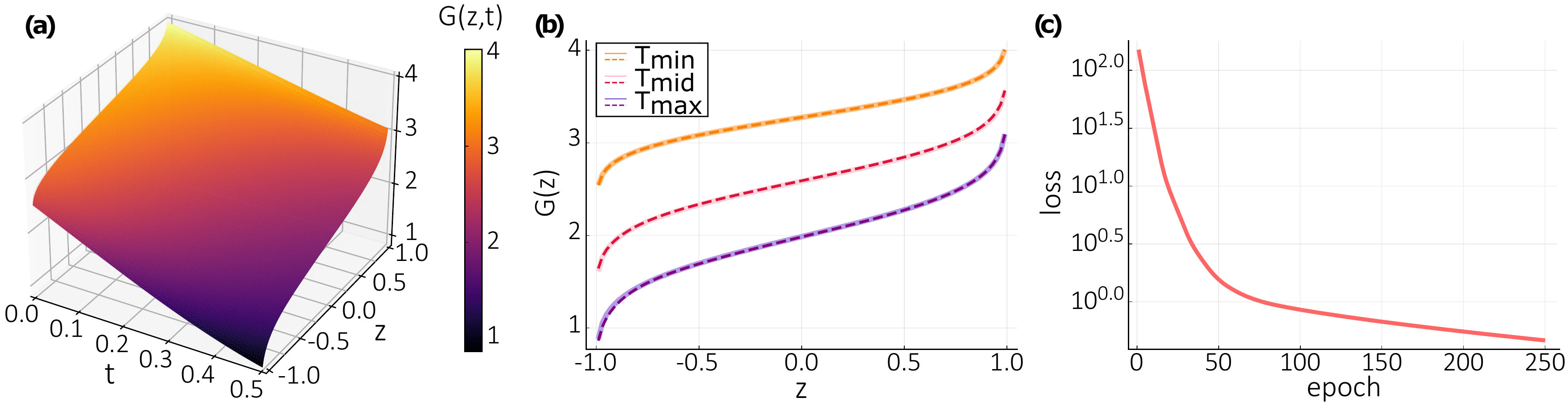}
\end{center}
\caption{\textbf{Training time evolution of Ornstein-Uhlenbeck process.} The results are shown for runs with analytic initial condition and parameters chosen as $\nu = 1, \sigma = 0.7, x_0 = 4$. \textbf{(a)} Surface plot for the trained DQC-based quantile function $G(z,t)$ that changes in time. \textbf{(b)} Slices of the quantum quantile function $G(z,t)$ shown at discrete time points $t = 0$ (labelled as $T_{\mathrm{min}}$ hereafter), $t = 0.25$ ($T_{\mathrm{mid}}$), and $t = 0.5$ ($T_\mathrm{max}$). \textbf{(c)} Loss as a function of epoch number showing the training progression and final loss for circuits shown in \textbf{(a, b)}.}
\label{fig:analytic_res}
\end{figure*}

The Ornstein-Uhlenbeck process is described by an SDE with an instantaneous diffusion term and linear drift. For a single variable process $X_t$ the OU SDE reads
\begin{align}
\label{eq:OU-SDE}
X_t = \nu (\mu - X_t) dt + \sigma dW_t ,
\end{align}
where the vector of underlying parameters $\bm{\xi} = (\nu, \mu, \sigma)$ are the speed of reversion $\nu$, the long-term mean level $\mu$, and the degree of volatility $\sigma$. The corresponding Fokker-Planck equation for the probability density function $p(x,t)$ reads
\begin{align}
\label{eq:OU-FP}
    \frac{\partial p(x,t)}{\partial t} = \nu \frac{\partial }{\partial x} \big( x p  \big) + \frac{\sigma^2}{2} \frac{\partial^2 p}{\partial x^2} .
\end{align}
When rewritten in the quantilized form, it becomes a PDE for the quantile mechanics,
\begin{align}
\label{eq:QE_OU}
    \frac{\partial Q(z,t)}{\partial t} = \nu \big[\mu - Q(z,t) \big] + \frac{\sigma^2}{2} \left( \frac{\partial Q}{\partial z} \right)^{-2} \frac{\partial^2 Q}{\partial z^2},
\end{align}
which follows directly from the generic Eq.~\eqref{eq:QE}. In the following we take the speed of reversion to be positive, $\nu > 0$ and adjust the long-term mean level to zero, $\mu = 0$.
\begin{figure}[b]
\centering
\includegraphics[width=1.0\linewidth]{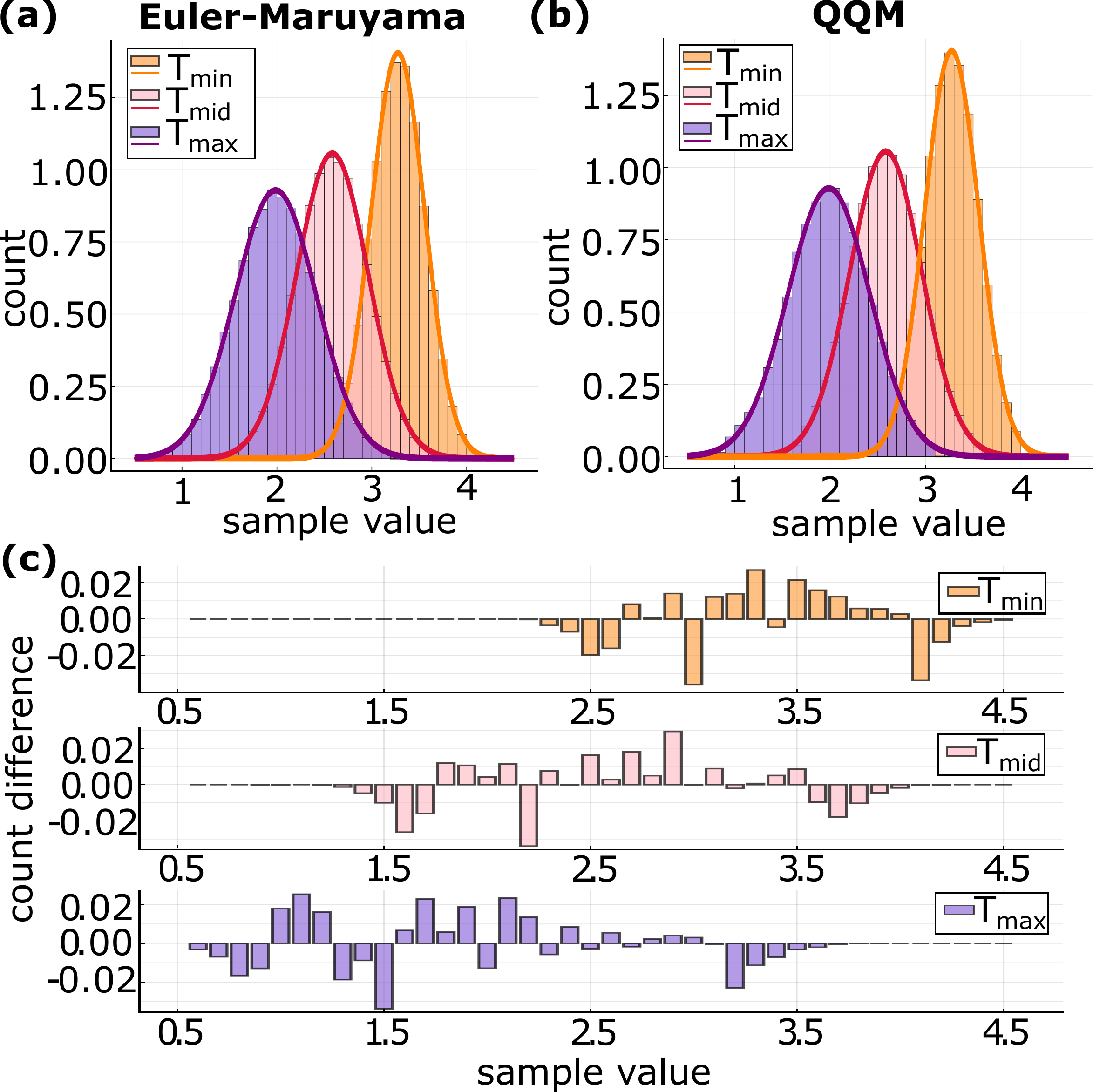}
\caption{\textbf{Comparison of histograms from the numerical SDE integration and QQM training.} \textbf{(a)} Distribution of samples from the Euler-Maruyama SDE solver (binned counts divided by the total number of samples $N_{\mathrm{s}}$), shown against the analytic PDF at three time points $T_{\mathrm{min}}$ ($t=0$), $T_{\mathrm{mid}}$ ($t = 0.25$), and $T_\mathrm{max}$ ($t = 0.5$). $N_{\mathrm{s}} = 100,000$ samples are taken, and parameters are the same ($\nu = 1$, $\sigma = 0.7$, $x_0 = 4$). \textbf{(b)}. Distribution of samples generated from the DQC-based quantile function, for the training described in Fig.~\ref{fig:analytic_res}. \textbf{(c)} Bar height difference between \textbf{(a)} and \textbf{(b)} shown for $T_{\mathrm{min}}$ (top), $T_{\mathrm{mid}}$ (middle), and $T_{\mathrm{max}}$ (bottom).}
\label{fig:analytic_hist}
\end{figure}

Having established the basics, we train the differentiable quantum circuit to match the OU QF. First, for the starting point of time, we train the circuit to represent a quantile function based on available data (see the workflow chart in Fig.~\ref{fig:workflow} and the discussion below).
Next, having access to the quantum QF at the starting point, we evolve it in time solving the equation
\begin{align}
\label{eq:QE2}
    \frac{\partial G(z,t)}{\partial t} = -\nu G(z,t) + \frac{\sigma^2}{2} \left( \frac{\partial G}{\partial z} \right)^{-2} \frac{\partial^2 G}{\partial z^2},
\end{align}
as required by QM [Eq.~\eqref{eq:QE_OU}]. This is the second training stage in the workflow chart shown in Fig.~\ref{fig:workflow}.

To check the results, we use the analytically derived PDF valid for the Dirac delta initial distribution $p(x, t_0) = \delta(x - x_0)$ peaked at $x_0$ that evolves as
\begin{align}
\label{eq:p_xt}
    p(x, t) = &\sqrt{\frac{\nu}{\pi \sigma^2 (1 - \exp[-2 \nu (t - t_0)])}} \times \\ \notag &\times \exp \left[ -\frac{\nu (x - x_0 \exp[-\nu (t -t_0)])^2}{\sigma^2 (1 - \exp[-2 \nu (t - t_0)])} \right],
\end{align}
and we can write the OU QF evolution as
\begin{align}
    \label{eq:Q_xt}
    Q(z,t) =& x_0 \exp\big[-\nu (t - t_0)\big] + \\ \notag &+ \sqrt{\frac{\sigma^2}{\nu} \Big(1 - \exp\big[-2 \nu (t-t_0)\big]\Big)} \hspace{1mm} \mathrm{inverf}(z),
\end{align}
where $\mathrm{inverf}(x)$ denotes the inverse error function~\cite{AbramowitzStegun}. This provides us a convenient benchmark of a simple case application and allows assessing the solution quality. Additionally we use Euler-Maruyama integration to compare results with the numerical sampling procedure with fixed number of shots.

\textit{Ornstein-Uhlenbeck with analytic initial condition.} To highlight the generative power of the QQM approach we start by simulating the OU evolution $G(z,t)$ starting from the known initial condition. This is set as $G(z,0) = Q(z,0)$ being the analytic solution [Eq.~\eqref{eq:Q_xt}] or can be supplied as a list of known samples associated to latent variable values. To observe a significant change in the statistics and challenge the training, we choose the dimensionless SDE parameters as $\nu = 1$, $\sigma = 0.7$, $x_0 = 4$, and $t_0 = -0.2$ such that we evolve a narrow normal distribution with strongly shifted mean into a broad normal distribution at $\mu = 0$. 
We use DQCs with $N=6$ qubits and a single cost operator being the total Z magnetization, $\hat{\mathcal{C}} = \sum_{j=1}^N \hat{Z}_j$. 
For simplicity we train the circuit using a uniformly discretized grid with $\mathcal{Z}$ containing $21$ points from $-1$ to $1$, and $\mathcal{T}$ containing $20$ values from $0.0$ to $0.5$. To encode the function we use the product-type feature maps~\cite{Mitarai2018,Kyriienko2021} chosen as $\hat{\mathcal{U}}_{\phi}(t) = \bigotimes_{j=1}^N \exp\big[-i \arcsin(t) \hat{Y}_j/2\big]$ and $\hat{\mathcal{U}}_{\phi'}(z) = \bigotimes_{j=1}^N \exp\big[-i \arcsin(z) \hat{X}_j/2 \big]$. The variational circuit corresponds to HEA with the depth of six layers of generic single-qubit rotations plus nearest-neighbor CNOTs. We exploit the floating boundary handling, and choose a mean squared error (MSE) as the distance measure, $\mathfrak{D}(a,b) = (a-b)^2$. The system is optimized for a fixed number of epochs using the $\texttt{Adam}$ optimizer for gradient-based training of variational parameters $\bm{\theta}$. We implement this with a full quantum state simulator in a noiseless setting. This is realized in \texttt{Yao.jl}~\cite{Luo2019} --- a \texttt{Julia} package that offers state-of-the-art performance.

We present the results of DQC training in Fig.~\ref{fig:analytic_res}. In Fig.~\ref{fig:analytic_res}(a) we show the trained quantum QF as a function of time $t$ and the latent variable $z$. Choosing three characteristic points of time $t = \{ 0.0, 0.25., 0.50\}$ that we label as $\{ T_{\mathrm{min}}, T_{\mathrm{mid}}, T_{\mathrm{max}} \}$, we plot the corresponding quantile functions at these times [Fig.~\ref{fig:analytic_res}]. The dashed curves from the DQC training closely follow ideal QFs shown by solid curves. Additionally, in Fig.~\ref{fig:analytic_res}(c) we show the training loss as a function of epoch number, noting a rapid convergence as the circuit is expressive enough to represent changes of initial QF at increasing time, and thus providing us with evolved $G(z,t)$. 

Next, we perform sampling and compare the histograms coming from the Euler-Maruyama integration of OU SDE~\cite{Kloeden1992,Rackauckas2017} and the QQM training presented above. The results are shown in Fig.~\ref{fig:analytic_hist} for the same parameters as Fig.~\ref{fig:analytic_res}. In Fig.~\ref{fig:analytic_hist}(a) we show the three time slices of Euler-Maruyama trajectories, built with $N_{\mathrm{s}} = 100,000$ samples to see distributions in full. The counts are binned and normalized by $N_{\mathrm{s}}$, and naturally show excellent correspondence with analytical results. The sampling from trained quantile is performed by drawing random $z \sim \mathtt{uniform(-1,1)}$ for the same number of samples. In Fig.~\ref{fig:analytic_hist}(b) we observe that QQM matches well the expected distributions. Importantly, the training correctly reproduces the widening of the distribution and the mean reversion, avoiding the mode collapse that hampers adversarial training~\cite{TongLi2020,Gui2020}. To further corroborate our findings, we plot the difference between two histograms (Euler-Maruyama and QQM) in Fig.~\ref{fig:analytic_hist}(c), and observe that the count difference remains low at different time points.
\begin{figure}[t]
\begin{center}
\includegraphics[width=1.0\linewidth]{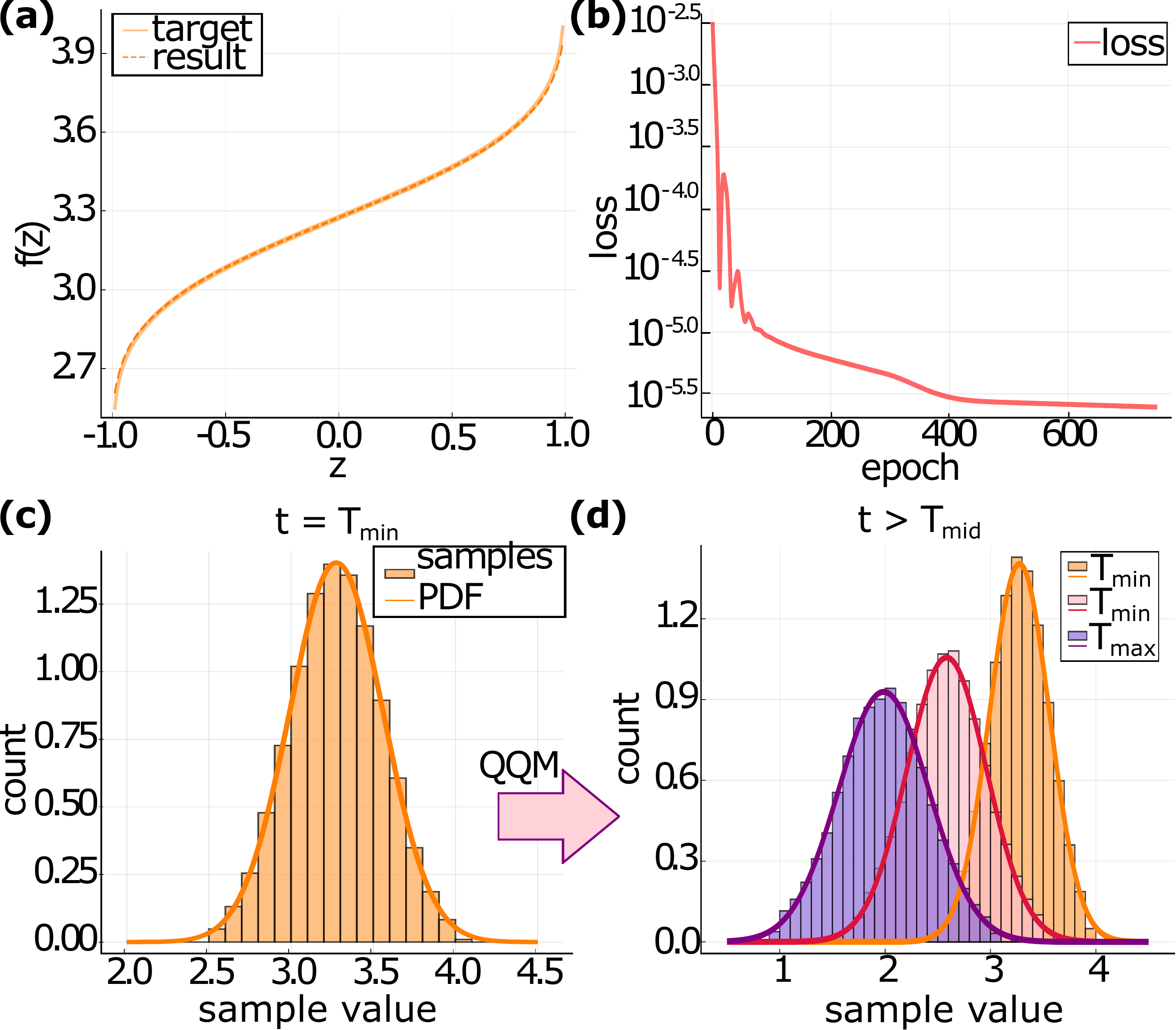}
\end{center}
\caption{\textbf{Quantile function trained on initial data.} \textbf{(a)} Trained QF for the Ornstein-Uhlenbeck process at $t=0$ (dashed curve labeled as \texttt{result}), plotted together with the known true quantile (solid line labeled as \texttt{target}, results overlay). The parameters are the same as for Fig.~\ref{fig:analytic_res}. \textbf{(b)} Training loss at different epochs, with the final epoch producing the QF in \textbf{(a)}. \textbf{(c)} Normalised histogram of samples from the data-trained QF, plotted against the analytic distribution (PDF). $N_{\mathrm{s}}=100,000$ random samples are drawn and bin counts are normalized by $N_{\mathrm{s}}$ as before. \textbf{(d)} Histograms for the data-trained QF evolved with quantum quantile mechanics, shown at three points of time.}
\label{fig:quantile_qcl}
\end{figure}

\textit{Ornstein-Uhlenbeck with data-inferred initial condition.} Next, we demonstrate the power of quantile function training from the available data (observations, measurements) corresponding to the Ornstein-Uhlenbeck process. Note that compared to the propagation of a known solution that is simplified by the boundary handling procedure, for this task we learn both the surface $G(z,t)$ and the initial quantile function $G(z,T_{\mathrm{min}})$. To learn the initial QF (same parameters as for Figs.~\ref{fig:analytic_res} and \ref{fig:analytic_hist}) we use QCL trained on the observations. The samples in the initial dataset are collected into bins and sorted in the ascending order as required by QF properties. From the original $N_{\mathrm{s}}=100,000$ that are ordered we obtain an interpolated curve. We get target values for QCL training choosing $N_\mathrm{points} = 43$ points in $\mathcal{Z}$ between $-1$ and $1$. We note that the training set is significantly reduced, and such data-frugal training holds as long as the QF structure is captured (monotonic increase). The training points are in the Chebyshev grid arrangement as $\cos[(2n-1) \pi /(2 N_\mathrm{points})]$ ($n = 1, 2, ..., N_{\mathrm{points}}$), this puts slight emphasis on training the distribution tails around $|z|\approx 1$. To make the feature map expressive enough that it captures full $z$-dependence for the trained initial QF, we use a tower-type product feature maps defined as $\hat{\mathcal{U}}_{\phi'}(z) = \bigotimes_{j=1}^N \exp\big[-i \arcsin(z) j \hat{Z}_j/2\big]$, where rotation angles depend on the qubit number $j$. For the training we again use a six-qubit register, and follow the same variational strategy as in the previous subsection. We observe that a high-quality solution with a loss of $\sim10^{-6}$ for $G(z,0)$ can be obtained at the number of epochs increased to few thousands, and we find that pre-training with product states allows reducing this number to hundreds with identical quality.

The results are shown in Fig.~\ref{fig:quantile_qcl}, with the QF trained from data shown in Fig.~\ref{fig:quantile_qcl}(a) by the dashed curve that overlays the target QF. The circuit converges to $10^{-6}$ loss level [Fig.~\ref{fig:quantile_qcl}(b)], being close to the model expressivity limit of the feature map itself.
Performing sample generation at the initial time, in Fig.~\ref{fig:quantile_qcl} we observe good correspondence with the expected PDF (sampling procedure is the same as in Fig.~\ref{fig:analytic_hist}). At the same time we note that small deviations in trained QF (and its derivative) lead to significant deviations for statistics, stressing the importance of expressive circuits and stable training.
Using the trained QF as the initial condition, we evolve the system as before (training details are presented in Supplemental Information). We perform generative modelling at later points of time $T_{\mathrm{mid}}$ and $T_{\mathrm{max}}$. The histograms in Fig.~\ref{fig:quantile_qcl} confirm the high quality of sampling, and show that the approach is suitable for time series generation.\vspace{1mm}



\noindent\textbf{qGAN-based generative modelling}

\noindent Generative adversarial networks represents one of the most successful strategies for generative modelling~\cite{Gui2020}. It is used in various areas, ranging from the fashion industry to finance, where GANs are used to enrich financial datasets. The latter is specially relevant when working with relatively scarce or sensitive data. The structure of GAN is represented by two neural networks: a generator $G_{\mathrm{NN}}$ and a discriminator $D_{\mathrm{NN}}$. The generator takes a random variable $z \sim p_z(z)$ from a latent probability distribution $p_z(z)$. This is typically chosen as a uniform (or normal) distribution for $z \in (-1,1)$. 
\begin{figure}[t]
\begin{center}
\includegraphics[width=1.0\linewidth]{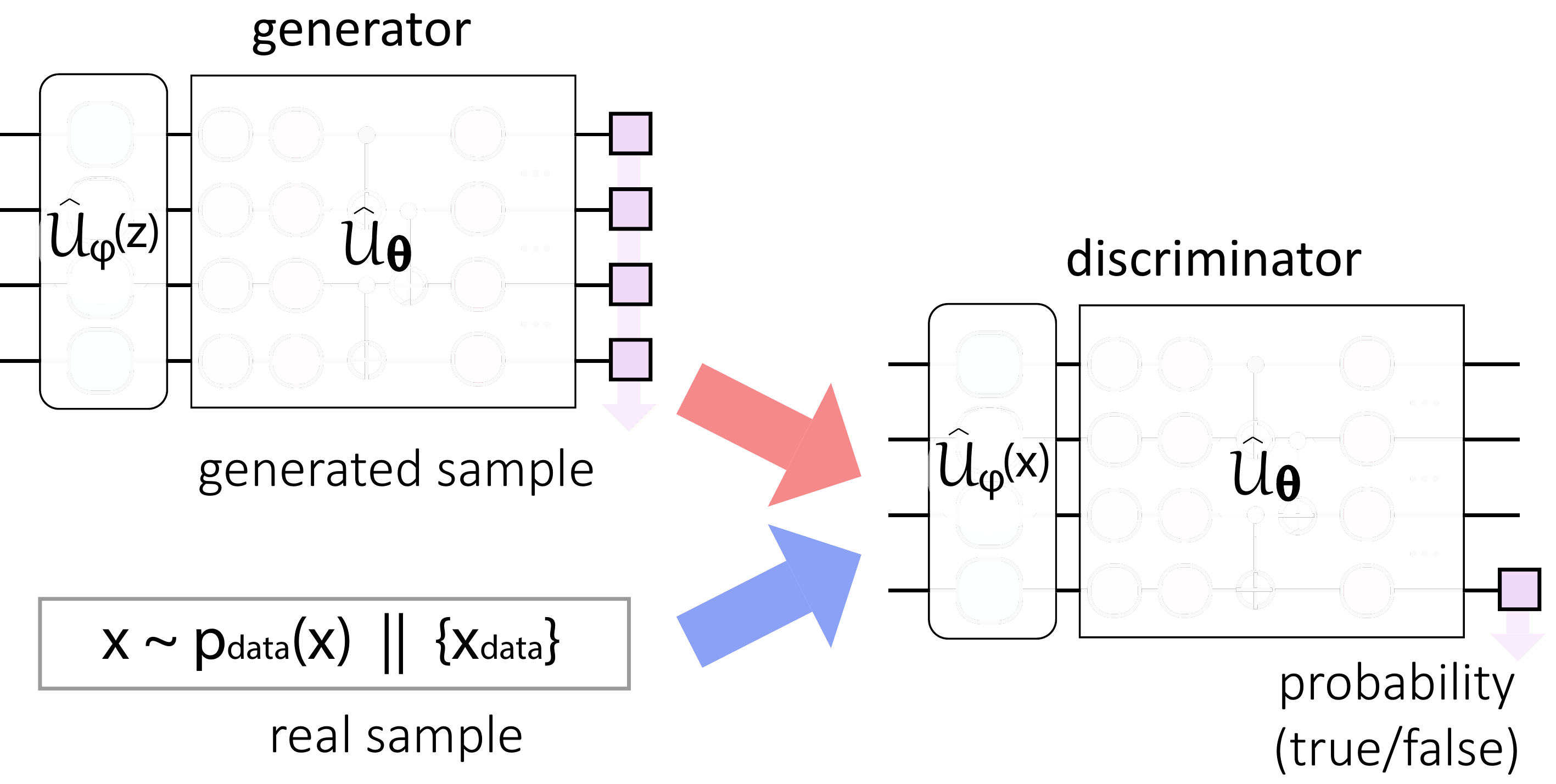}
\end{center}
\caption{\textbf{Quantum GAN workflow.} The quantum circuits are used both for generative modelling at $t = T_{\mathrm{min}}$ (generator) and discrimination between real and fake samples (discriminator). The generator circuit $G_{\mathrm{Q}}(z)$ is composed of the product feature map and HEA variational circuit. The discriminator $D_{\mathrm{Q}}(x)$ is trained to distinguish samples from the initial data distribution.}
\label{fig:qGAN_circuit}
\end{figure}
Using a composition $g_L \circ ... \circ g_2 \circ g_1(z)$ of (nonlinear) operations $\{ g_i \}_{i=1}^L$ such that the generator prepares a fake sample $G_{\mathrm{NN}}(z)$ from the generator's probability distribution $p_G$, $G_{\mathrm{NN}}(z) \sim p_G(G_{\mathrm{NN}}(z))$. The goal is to make samples $\{ G_{\mathrm{NN}}(z) \}_{s=1}^{N_\mathrm{s}}$, as close to the training dataset as possible, in terms of their sample distributions. If true samples $x \sim p_{\mathrm{data}}(x)$ are drawn from a (generally unknown) probability distribution $p_{\mathrm{data}}(x)$, our goal is to match $p_G(G_{\mathrm{NN}}(z)) \approx p_{\mathrm{data}}(x)$. This is achieved by training the discriminator network $D_{\mathrm{NN}}$ to distinguish true from fake samples, while improving the quality of generated samples $\{ G_{\mathrm{NN}}(z) \}_{s=1}^{N_\mathrm{s}}$, optimizing a minimax loss
\begin{align}
\label{eq:minimax_loss}
    &\min\limits_{G_{\mathrm{NN}}} \max\limits_{D_{\mathrm{NN}}} \mathcal{L}_{\mathrm{GAN}} = \min\limits_{G_{\mathrm{NN}}} \max\limits_{D_{\mathrm{NN}}} \bigg\{ \mathbb{E}_{x \sim p_{\mathrm{data}}(x)} \big[\log D_{\mathrm{NN}}(x)\big] \\ \notag &+ \mathbb{E}_{z \sim p_{z}(z)} \big[\log (1 - D_{\mathrm{NN}}(G_{\mathrm{NN}}(z))) \big] \bigg\},
\end{align}
where $D_{\mathrm{NN}}$ and $G_{\mathrm{NN}}$ are the trainable functions represented by the discriminator and generator, respectively. The first loss term in Eq.~\eqref{eq:minimax_loss} represents the log-likelihood maximization that takes a \emph{true} sample from the available dataset, and maximizes the probability for producing these samples by adjusting variational parameters. The second term trains $G_{\mathrm{NN}}$ to minimize the chance of being caught by the discriminator. 
Most importantly, we note that $G_{\mathrm{NN}}(z)$ is a function that converts a random sample $z \sim \mathtt{uniform(-1,1)}$ into a sample from the trained GAN distribution --- therefore representing a \emph{quantile-like function}. This is the connection we develop further using the qGAN training.
\begin{figure}[t]
\begin{center}
\includegraphics[width = 1.0\linewidth]{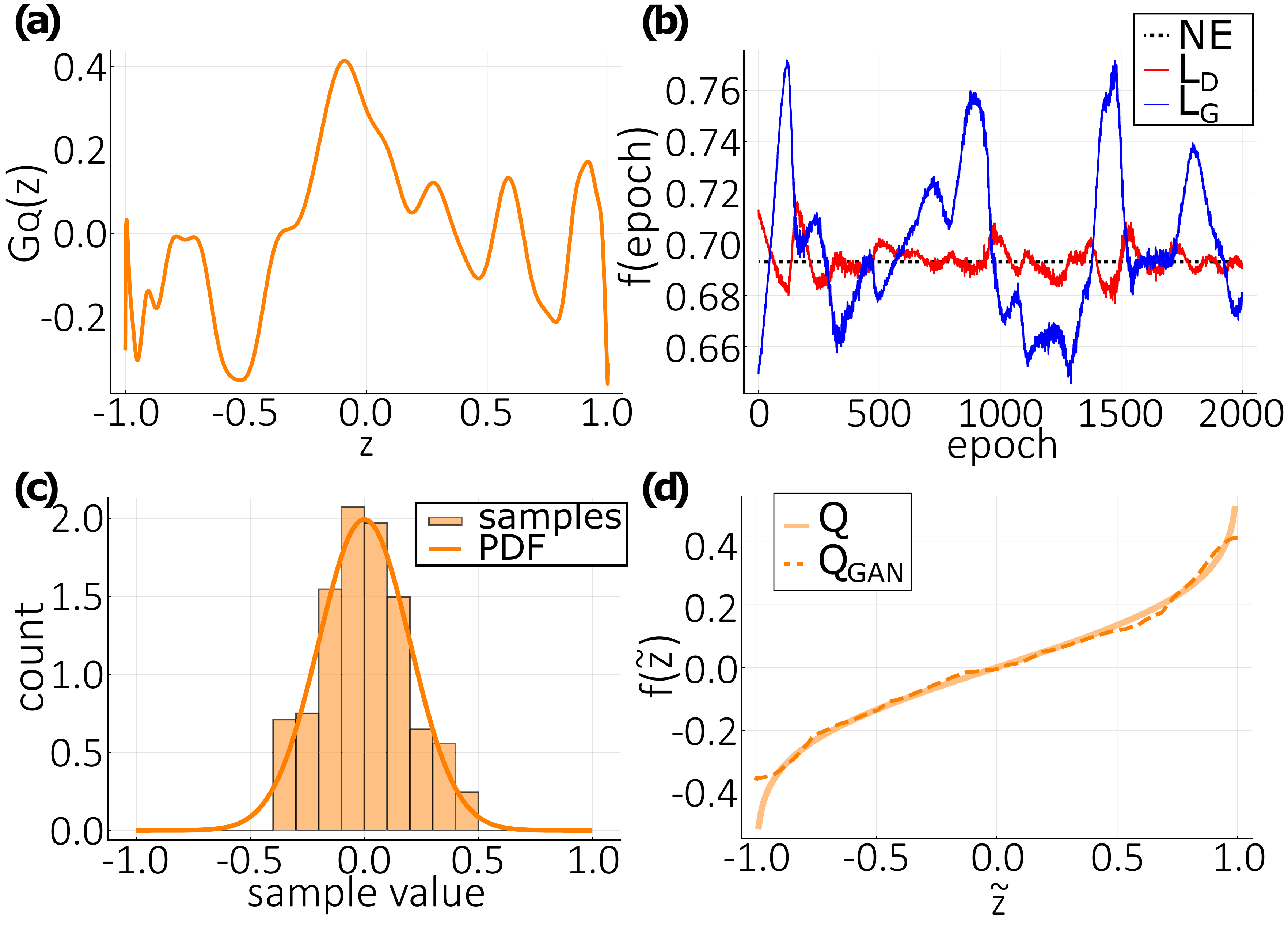} 
\end{center}
\caption{\textbf{qGAN training and fixed-time sampling.} \textbf{(a)} Generator function is shown for the optimal training angles. \textbf{(b)} Generator ($L_G$, red) and discriminator ($L_D$, blue) loss terms at different epochs. The Nash equilibrium at $-\ln(1/2)$ is shown by black dotted line (NE). \textbf{(c)} Normalized histogram for qGAN sampling ($N_{\mathrm{s}}=100,000$), as compared to the target normal distribution ($\mu = 0$, $\sigma = 0.2$). \textbf{(d)} Ordered quantile $Q_{\mathrm{GAN}}(\tilde{z})$ from the resulting qGAN generator shown in \textbf{(a)} (dashed curve), as compared to the true QF of the target distribution (solid curve).}
\label{fig:qgan_training_fig}
\end{figure}

Quantum GANs follow the same ideology as their classical counterparts, but substitute the neural representation of the generator $G_{\mathrm{NN}}$ and/or discriminator $D_{\mathrm{NN}}$ by quantum neural networks. In the following, we denote them as $G_{\mathrm{Q}}$ and $D_{\mathrm{Q}}$, respectively. The schedule of qGAN training and circuits are presented in Fig.~\ref{fig:qGAN_circuit}. To apply qGANs for the same task of OU process learning, we concentrate on a continuous qGAN that uses the feature map encoding~\cite{Romero2021,Anand2021}. We follow the training strategy from Ref.~\cite{Romero2021}. We try to model the normal distribution with zero mean and standard deviation of $0.2$. Both the discriminator and generator use $N=6$ registers with the expressive Chebyshev tower feature map~\cite{Kyriienko2021} followed by $d=6$ HEA ansatz. The readout for the generator uses $\langle \hat{Z}_1 \rangle$ expectation, and the discriminator has the cost function such that we readout $(\langle \hat{Z}_1 \rangle + 1)/2 \in [0,1]$ modeling the probability. As before, we use \texttt{Adam} and train the qGAN for $2000$ epochs using the loss function \eqref{eq:minimax_loss}. Due to the minimax nature of the training, the loss oscillates and instead of reaching (global) optimum qGAN tries to reach the Nash equilibrium. Unlike QQM training, we cannot simply use variational parameters for the final epoch, and instead test the quality throughout. To get the highest quality generator we test how close together are the discriminator ($L_\mathrm{D}$) and generator ($L_\mathrm{G}$) loss terms. If they are within $\epsilon = 0.1$ distance we perform the Kolmogorov-Smirnov (KS) test~\cite{Zoufal2019} and check the distance between the currently generated samples and the training dataset. The result with minimal KS is chosen. We stress that KS is not used for training, and is exclusively for choosing the best result.

The results for qGAN training are shown in Fig.~\ref{fig:qgan_training_fig}. The trained generator $G_{\mathrm{Q}}(z)$ is shown as a function of the latent variable $z$. We note that it has a strongly-oscillating nature. The training is shown separately for the generator (blue curve) and discriminator (red curve) loss terms. They oscillate around the analytic value for the Nash equilibrium (black dotted line, \texttt{NE}), and briefly settle around NE after $1600$ epochs where optimal circuit parameters are saved. We sample the qGAN generator using $N_{\mathrm{s}}=100,000$ and plot the normalized histogram in Fig.~\ref{fig:qgan_training_fig}(c). We observe that the distribution roughly matches the target (solid curve, PDF), though finer points are missing [cf. Fig.~\ref{fig:quantile_qcl}(c)], including the missing tail at negative values. Naturally, the generator of qGAN $G_{\mathrm{Q}}(z)$ does the same job as the trained quantile function $G(z)$ from previous subsections. We proceed to connect the two explicitly.
\begin{figure}[t]
\begin{center}
\includegraphics[width = 1.0\linewidth]{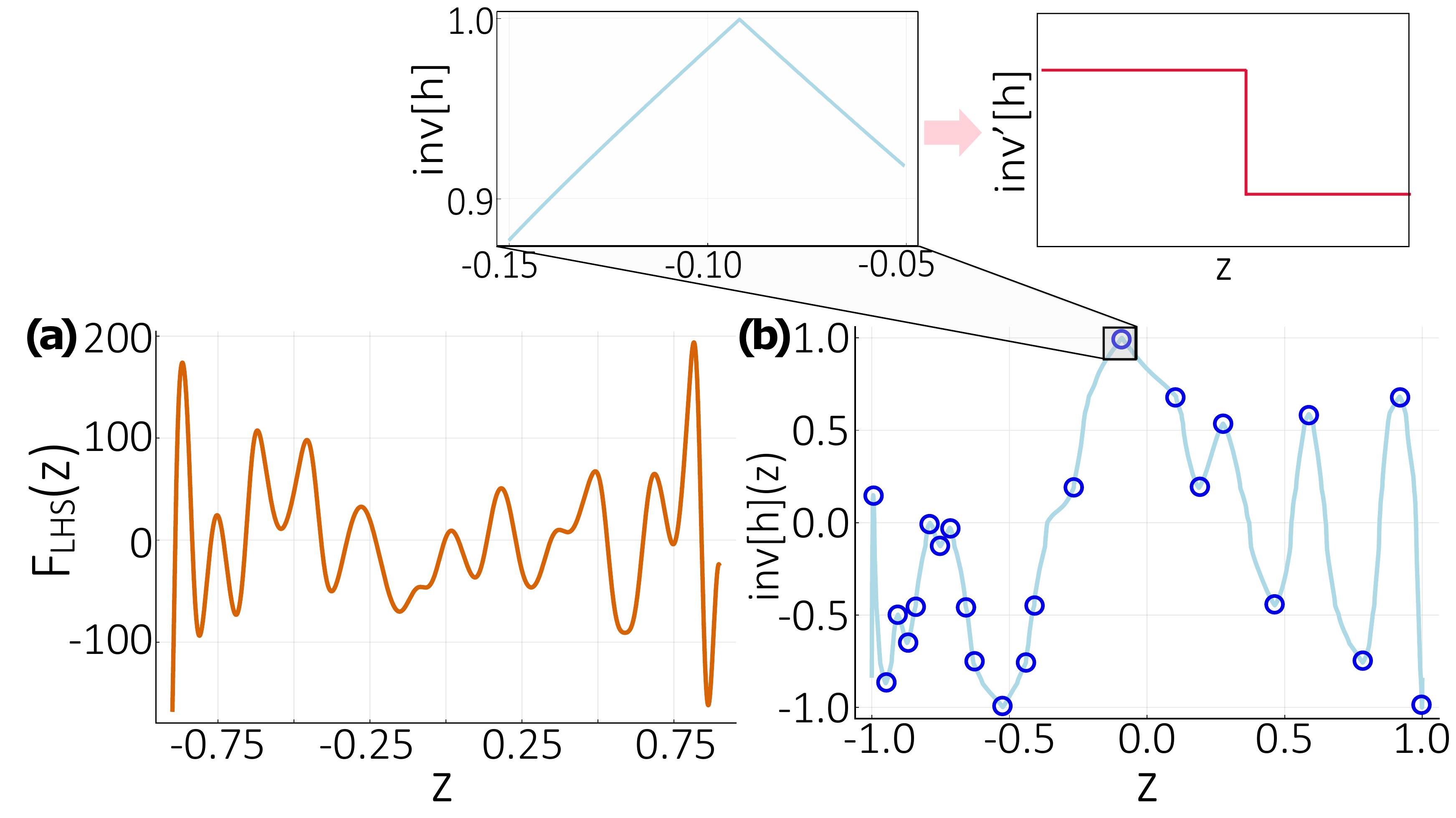} 
\end{center}
\caption{\textbf{qGAN quantile function analysis.} In \textbf{(a)} we plot the LHS of Eq.~\eqref{eq:chain_quantile_ode} for the trained qGAN generator $G_{\mathrm{Q}}(z)$ shown in Fig.~\ref{fig:qgan_training_fig}(a). The resulting function is oscillatoric but smooth. \textbf{(b)} Inverse mapping function $\mathrm{inv}[h]$ shown as a function of $z$. It transforms $G_{\mathrm{Q}}(z)$ into the increasing quantile function $Q_{\mathrm{GAN}}(\tilde{z})$ that is plotted in Fig.~\ref{fig:qgan_training_fig}(d). We highlight the non-differentiable points by blue circles, and zoom in on the characteristic behavior in the inset above. The derivative is not defined at the discontinuity (top right inset).}
\label{fig:qgan_qode_fig}
\end{figure}

\textit{Reordered quantile functions and their DEs.} The main difference between the quantile function and the generator of qGAN is that the true QF is a strictly monotonically increasing function, while the qGAN generator $G_{\mathrm{Q}}$ is not. We can connect them by noticing that qGAN works with the latent variable $z \in \mathcal{Z}$, which we can rearrange into a QF by ordering the observations and assigning them the \emph{ordered} latent variable $\tilde{z} \in \tilde{\mathcal{Z}}$ (both functions produce the same sample distribution). It is convenient to define a mapping $h: \tilde{\mathcal{Z}} \rightarrow \mathcal{Z}$ for $G_{\mathrm{Q}}$ which rearranges it into increasing form, $Q_{\mathrm{GAN}}(\tilde{z}) = G_{\mathrm{Q}}(h(\tilde{z}))$. In practice, finding $h(\tilde{z})$ requires the evaluation of $G_{\mathrm{Q}}(z)$ $\forall$ $z \in \mathcal{Z}$ and re-assigning the samples to values of $\tilde{\mathcal{Z}}$ in ascending order. Importantly, both $h$ and its inverse $\mathrm{inv}[h]: \mathcal{Z} \rightarrow \tilde{\mathcal{Z}}$ can be defined in this process.
In Fig.~\ref{fig:qgan_training_fig}(d) we show the results of reordering for the generator function in Fig.~\ref{fig:qgan_training_fig}(a). The reordered quantile $Q_{\mathrm{GAN}}$ is plotted (dashed curve), approximately matching the target quantile (solid curve). We observe that the center of the quantile is relatively well approximated but the tails are not (particularly for $\tilde{z}<0$). This agrees with what is observed in the sampling shown in Fig.~\ref{fig:qgan_training_fig}(c). Having established the correspondence for qGAN-based generative modeling and quantile-based modeling we ask the question: can we apply differential equations to the quantile-like function to add differential constraints, and evolve the system in time enabling generative modelling?



\begin{flushleft}\textbf{DISCUSSION}\end{flushleft}\vspace{-2mm}
The answer to the question above is far from trivial. To use a re-ordered qGAN quantile function for further training and time-series generation we need to account for the mapping when writing differential equations of quantile mechanics. 
Let us look into a specific case to develop an intuition on the behavior of re-ordered quantile functions with differential equations. A quantile function $Q(\tilde{z})$ of a normal distribution with mean $\mu$ and standard deviation $\sigma$ satisfies a quantile ODE~\cite{Shaw2008,CarilloToscani}
\begin{align}
    \frac{d^2 Q}{d\tilde{z}^2} - \frac{Q-\mu}{\sigma^2}\left(\frac{dQ}{d\tilde{z}}\right)^2 = 0,
    \label{eq:qode_normal}
\end{align}
where we use the tilde notation $\tilde{z}$ to highlight that this is an ordered variable.
Assuming perfect training such that $Q_{\mathrm{GAN}}(\tilde{z}) = G_{\mathrm{Q}}(h(\tilde{z}))$ closely matches $Q(\tilde{z})$, we substitute it into Eq.~\eqref{eq:qode_normal}, and observe that the original qGAN generator obeys
\begin{align}
    \frac{d^2 G_{\mathrm{Q}}(z)}{dz^2} - \frac{G_{\mathrm{Q}}(z)-\mu}{\sigma^2}\left(\frac{dG_{\mathrm{Q}}(z)}{dz}\right)^2 = \frac{\mathrm{inv}[h]''(z)} {\mathrm{inv}[h]'(z)} \frac{dG_{\mathrm{Q}}(z)}{dz}.
    \label{eq:chain_quantile_ode}
\end{align}
The left-hand side (LHS) of Eq.~\eqref{eq:chain_quantile_ode} has the same form as for the true QF [cf. Eq~\eqref{eq:qode_normal}], but the right-hand side (RHS) differs from zero and involves derivatives of the inverted mapping function $\mathrm{inv}[h](z)$. This has important implications for training $G_{\mathrm{Q}}(z)$ with differential constraints, as the loss term includes the difference between LHS and RHS. Let us analyze the example of the quantile ODE in Eq.~\eqref{eq:chain_quantile_ode}. In Fig.~\ref{fig:qgan_training_fig}(a) we plot the LHS for $G_{\mathrm{Q}}(z)$ coming from the qGAN training. The result is a smooth function, and we expect all relevant terms, including derivatives $dG_{\mathrm{Q}}(z)/dz$ or $d^{2}G_{\mathrm{Q}}(z)/dz^2$, can be evaluated and trained at all points of the latent space. However, the problem arises when RHS enters the picture. The additional term strongly depends on the contributions coming from inverse map derivatives $\mathrm{inv}[h]'$ and $\mathrm{inv}[h]''$. At the same time we find that the map from a non-monotonic to a monotonically increasing function is based on a multivalued function (see discussion and examples in Supplemental Information). Furthermore the inverse of the map (along with the map itself) is continuous but not smooth --- it becomes non-differentiable at some points due to $G_{\mathrm{Q}}(z)$ oscillations. As an example in Fig.~\ref{fig:qgan_qode_fig}(b) we show $\mathrm{inv}[h]$ from training in Fig. \ref{fig:qgan_training_fig}, highlighting the points with non-analytic behavior blue circles. The inset for Fig.~\ref{fig:qgan_qode_fig}(b) clearly shows the discontinuity. This translates to the absence of $\mathrm{inv}[h]'(z)$ at a set of points, which unlike zero derivatives cannot be removed by reshuffling the terms in the loss function. The discovered unlikely property of the mapping puts in jeopardy the attempts to use differential-based learning for qGAN generators. While more studies are needed to estimate the severity of discontinuities (and if the set of such points can be excluded to yield stable training), our interim conclusion is that quantile functions in the canonical increasing form are more suitable for evolution and time series generation.\vspace{1mm}

\textit{Closing notes.} We proposed a distinct quantum algorithm for generative modelling from stochastic differential equations. Summarizing the findings, we have developed the understanding of generative modelling from stochastic differential equations based on the concept of quantile functions. We proposed to represent the quantile function with a trainable (neural) representation, which may be classical- or quantum-based. In particular, we focused on parameterizing the trainable quantile function with a differentiable quantum circuit that can learn from data and evolve in time as governed by quantile mechanics equations. Using Ornstein-Uhlenbeck as an example, we benchmark our approach and show that it gives a robust strategy for generative modelling in the NISQ setting. Furthermore, we notice that adversarial schemes as continuous qGAN lead to modified quantile-like function that potentially have intrinsic obstacles for evolving them in time. 
We conclude by saying that the strategy we propose uses the large expressive power of quantum neural networks, and we expect elements of the approach can be used for other architectures.

\textit{Ethics declaration.} A patent application for the method described in this manuscript has been submitted by Qu\&Co with OK, AEP and VEE as inventors.



%

\begin{flushleft}\textbf{METHODS}\end{flushleft}\vspace{-2mm}

\noindent \small{In this section we describe the details of circuit differentiation and the proposed boundary handling procedure for multivariate functions.\vspace{1mm}

\noindent\textbf{Calculating second-order derivatives}

\noindent For solving SDEs we need to access the derivatives of the circuits representing quantile functions $dG/dz$, $d^2 G/dz^2$, where $z$ is a latent variable. This can be done using automatic differentiation techniques for quantum circuits, where for near-term devices and specific gates we can use the parameter shift rule~\cite{Mitarai2018,Schuld2019,Mitarai2020}. The differentiation of quantum feature maps follows the DQC strategy~\cite{Kyriienko2021}, where we estimate $d G/dz$ as a sum of expectation values 
\begin{align}
    \frac{dG(z)}{dz} = \frac{1}{2} \sum_{j=1}^N \varphi_j'(z) \Big(\langle G^+_j\rangle - \langle G^-_j \rangle \Big)
\end{align}
where $\langle G^+_j\rangle$ and $\langle G^-_j \rangle$ denote the evaluation of the circuit with the $j$-th gate parameter shifted positively and negatively by $\pi/2$. This generally requires $2N$ circuit evaluations.  The (non-linear) function $\varphi_j(z)$ represents the $z$-dependent rotation phase for $j$-th qubit. A popular choice is $\varphi_j(z) = \arcsin(z)$ (same for all qubits) referred as a product feature map, and other choices include tower feature maps~\cite{Kyriienko2021}.

Quantum circuit differentiation for higher-order derivatives was recently considered in several studies~\cite{Mitarai2020,Cerezo2021b,Mari2021}. Extending the feature map differentiation to the second order, we use parameter shift rule alongside the product rule, and calculate $d^2 G/dz^2$ as
\begin{align}
\label{eq:2nd_derivative}
    &\frac{d^2G(z)}{dz^2} = \frac{1}{2} \sum_{j=1}^N \varphi_j''(z) \Big(\langle G^+_j\rangle - \langle G^-_j \rangle \Big) \\ \notag &+ \frac{1}{4} \sum_{j=1}^N \sum_{k=1}^N \varphi_j'(z) \varphi_k'(z) \Big(\langle G^{++}_{jk}\rangle - \langle G^{+-}_{jk} \rangle - \langle G^{-+}_{jk}\rangle + \langle G^{--}_{jk} \rangle\Big),
\end{align}
where $\langle G^{++}_{jk}\rangle$ ($\langle G^{--}_{jk}\rangle$) denotes the evaluation of the circuit with the rotation angles on qubits $j$ and $k$ are shifted positively (negatively) by $\pi/2$. Similarly, $\langle G^{+-}_{jk}\rangle$ ($\langle G^{-+}_{jk}\rangle$) are defined for shifts in opposite directions. If implemented naively, the derivative in Eq.~\eqref{eq:2nd_derivative} requires $2N + 4N^2$ evaluations of circuit expectation values. This can be reduced by making use of symmetries in the shifted expectation values and reusing/caching previously calculated values, i.e. those for the function and its first-order derivative. The reduced number of \textit{additional} circuit evaluations required for the calculation of the second derivative is $2N^2$.\vspace{1mm}

\noindent \textbf{Boundary handling for PDEs} 

\noindent As we consider differential equations with more than one independent variable, we need to develop a strategy for implementing (handling) the boundary in this situation. Let us consider a function of $n$ variables. We consider an initial condition of $f(t=0, \bm{z}) = u_0(\bm{z})$, where $\bm{z}$ is a vector of $n-1$ independent variables, and the first variable usually corresponds to time. Here we extend several techniques, corresponding to pinned type and floating type boundary handling, previously considered for the single-variable case in Ref.~\cite{Kyriienko2021}.

When considering just one independent variable, a pinned boundary handling corresponds to encoding the function as $f(t) = G(t)$. The boundary is then `pinned' into place by use of a boundary term in the loss function $\mathcal{L}^B = \Big[f(t_0) - u_0 \Big]^2$. For multiple independent variables this generalizes to $f(t, \bm{z}) = G(t, \bm{z})$ and $\mathcal{L}^B = \sum_i \Big[f(t_0, \bm{z}_i)- u(\bm{z}_i) \Big]^2$, where $\{ \bm{z}_i \}$ are the set of points along $t=0$ at which the boundary is being pinned.

When using the floating boundary handling the boundary is implemented during the function encoding. In this case the function encoding is generalized from its single-variable representation, $f(t) = u_0 - G(0) + G(t)$, to the multivariate case as
\begin{align}
    f(t, \bm{z}) = u_0(\bm{z}) - G(0, \bm{z}) + G(t, \bm{z}).
\end{align}
This approach does not require the circuit-embedded boundary, but instead needs derivatives of $u_0(\bm{z})$ for calculating the derivatives of $f$ with respect to any $z \in \mathcal{Z}$.\vspace{2mm}
}

\begin{flushleft}\textbf{DATA AVAILABILITY}\end{flushleft}\vspace{-2mm}
\noindent \small{The data that support the findings of this study are available from the corresponding author upon reasonable request.}\vspace{1mm}


\begin{flushleft}\textbf{AUTHOR CONTRIBUTIONS}\end{flushleft}\vspace{-2mm}
\noindent \small{A.\,E.\,P. performed the calculations, described the results, and analyzed the qGAN performance. O.\,K. proposed the original idea and overseen the project together with V.\,E.\,E. All authors contributed to writing the manuscript and analyzing the results.}\vspace{1mm}

\begin{flushleft}\textbf{COMPETING INTERESTS}\end{flushleft}\vspace{-2mm}
\noindent \small{The authors declare that there are no competing interests.}\vspace{1mm}

\clearpage

\newpage

\begin{centering}
\Large{Supplemental Information}
\end{centering}
\vspace{2mm}

\normalsize{
\noindent \textbf{Analytic quantile reordering for qGAN generators}
}

\noindent To understand the reordering procedure for qGAN generators and strictly increasing quantile functions, let us consider a simple example. 
\begin{figure}[h]
\begin{center}
\includegraphics[width = 1.0\linewidth]{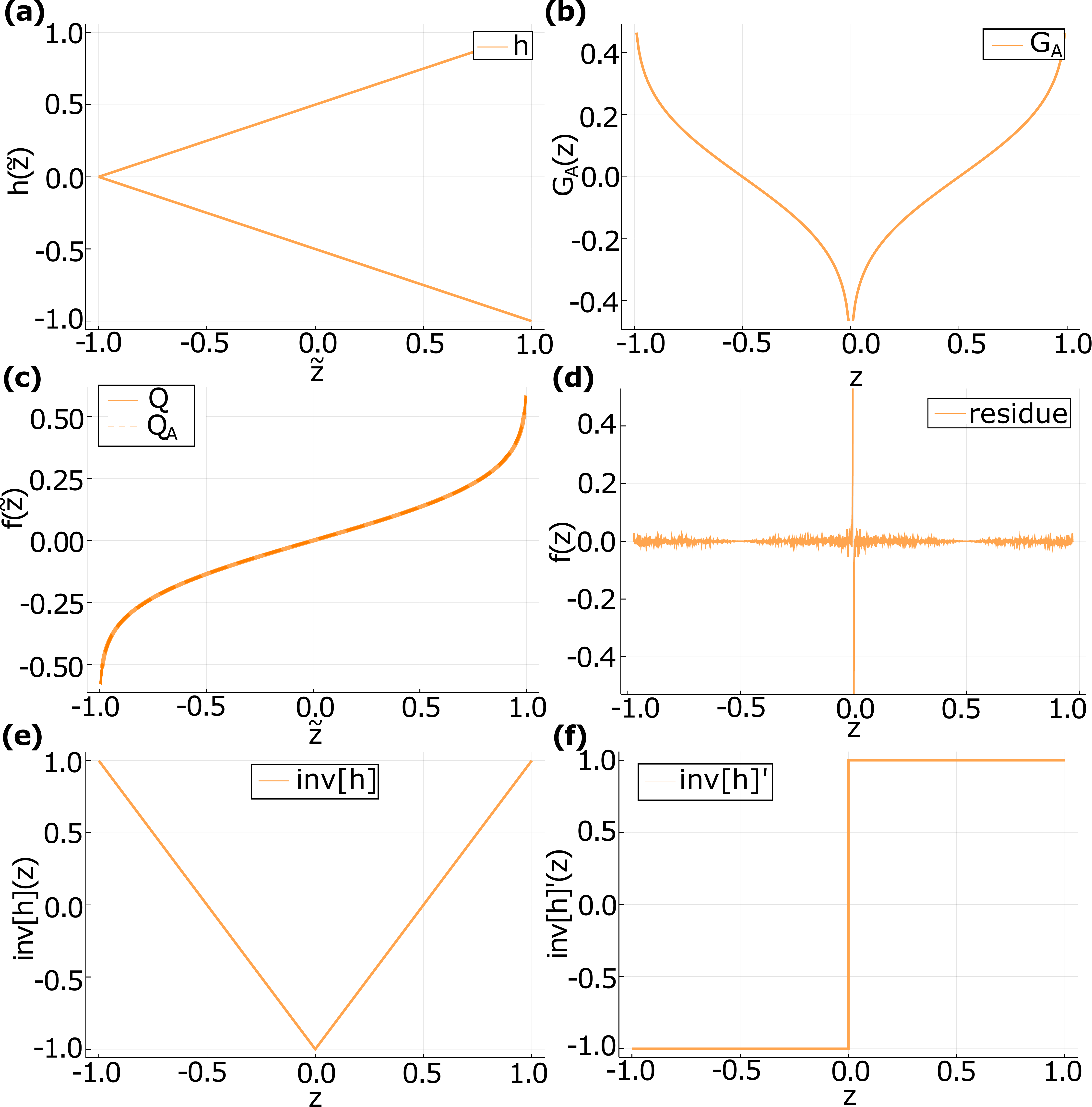} 
\end{center}
\caption{\textbf{qGAN quantile reordering example.} \textbf{(a)} Mapping function $h(\tilde{z}) = \pm (\tilde{z}+1)/2$. \textbf{(b)} Non-ordered quantile function $G_{\mathrm{A}}(z) = Q(\mathrm{inv}[h](z))$ corresponding to mapping in \textbf{(a)} and normal quantile function. \textbf{(c)} Known quantile function (solid curve, $Q$) compared to numerical ordering of $G_A$ (dashed line, $Q_A$). \textbf{(d)} RHS of Eq.~\eqref{eq:2nd_derivative} in the main text, plotted for $G_{\mathrm{A}}$ shown in \textbf{(b)}. Derivatives of $\mathrm{inv}[h]$ are calculated using finite differencing. \textbf{(e)} Inverse mapping function $\mathrm{inv}[h]$ plotted for different values of the latent variable $z$. We note the point of non-differentiability at $z = 0$. \textbf{(f)} Analytic first derivative of $\mathrm{inv}[h]$ that has a discontinuity at $z =0$.}
\label{fig:qode}
\end{figure}
\begin{figure*}[ht]
\begin{center}
\includegraphics[width=1.0\linewidth]{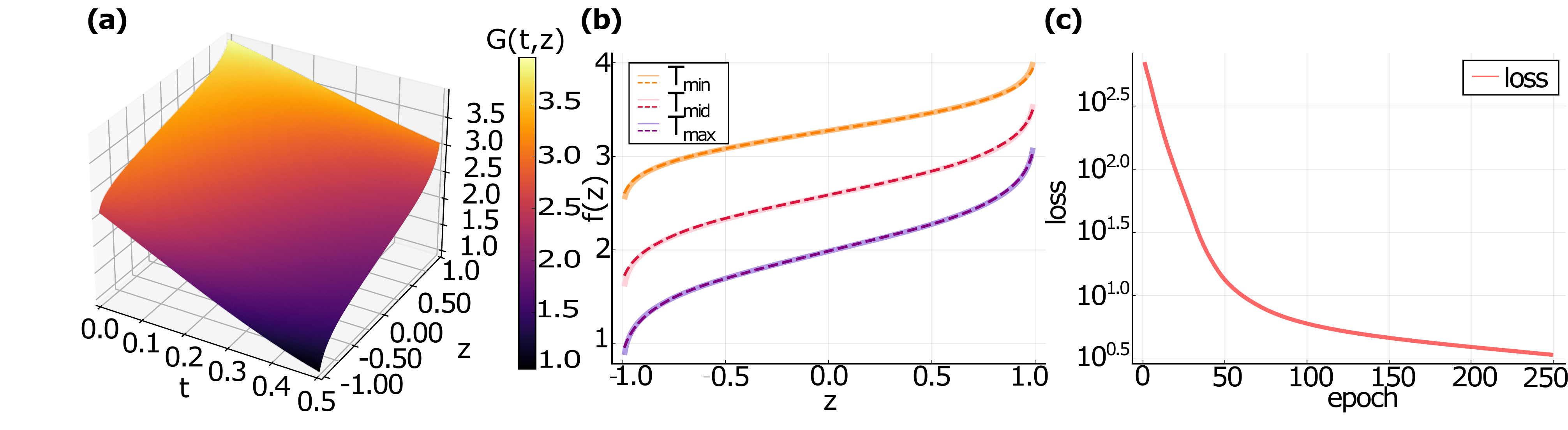}
\end{center}
\caption{\textbf{Time-evolution of Ornstein-Uhlenbeck process with data-inferred initial condition.} The initial conditions is taken from Fig.~\ref{fig:quantile_qcl} and we use the same OU model parameters. \textbf{(a)} Surface plot for the quantile function evolved in time. \textbf{(b)} Quantile functions shown at three time points (being the same as in the main text). \textbf{(c)} Loss as a function of epoch from training in \textbf{(a)} and \textbf{(b)}.}
\label{fig:obs_res}
\end{figure*}

We start by considering a quantile-like generator $G_{\mathrm{A}}(z)$ and use the mapping $h$ that reorders it into ideal quantile function (QF) for the normal distribution. The mapping reads $h(\tilde{z}) = \pm \left(\tilde{z}+1\right)/2$, and is shown in Fig.~\ref{fig:qode}(a) as a multivalued function. It ensures that if we start from the normal QF, we arrive to $G_{\mathrm{A}}(z)$ with a single dip. The corresponding qGAN-like generator $G_{\mathrm{A}}(z)$ with a single dip is shown in Fig.~\ref{fig:qode}(b) (we consider $\mu = 0$ and $\sigma = 0.2$). Our motivation is to understand how the presence of nonmonotonicity changes the behaviour of the system.

First, let us check that the reordering into increasing quantile function works as expected. Assigning the values of $G_{\mathrm{A}}(z)$ in the ascending order we get the reordered QF $Q_{\mathrm{A}}(\tilde{z})$. This is plotted in Fig.~\ref{fig:qode}(c), matching the ideal $Q(\tilde{z})$ as expected.
Once we have established the mapping for ideal re-ordering, let us look at its properties.
In the results section we discussed how the reordered quantile function from qGAN training matches the appropriate quantile ODE. We can perform the same check for the simple re-ordering presented above. Evaluating the difference between the RHS and LHS of Eq.~\eqref{eq:chain_quantile_ode} (akin to loss term) for $G_{\mathrm{A}}(z)$ and known mapping $h(\tilde{z})$, we observe that the difference remains zero everywhere (we have a perfect solution), apart from the middle point $z=0$ where it diverges [Fig.~\ref{fig:qode}(d)]. 
For calculating the derivatives of $\mathrm{inv}[h]$ we use finite differencing (Euler's method), similarly for the qGAN case, thus observing small noise coming from numerical differentiation. The reason behind the unfavorable loss term behavior can be tracked to the properties of the mapping function. We show the inverse mapping $\mathrm{inv}[h]$ plotted in Fig.~\ref{fig:qode}(e), and its derivative $\mathrm{inv}[h]'$ is presented in Fig.~\ref{fig:qode}(f). We see that $\mathrm{inv}[h]$ has a point of non-differentiability at $z=0$ and $\mathrm{inv}[h]'$ is discontinuous there. This provides the intuition behind the divergence. When training the DE-based loss for Eq.~\eqref{eq:chain_quantile_ode} with $z=0$ included the loss becomes non-trainable. We stress that the same is observed for the non-ideal $G_{\mathrm{Q}}$, where multiple non-differentiable points appear that we do not know in advance. The issue of developing efficient workflow for training $G_{\mathrm{Q}}(z)$, also including the time dimension, remains an important area for the future research.\vspace{1mm}

\noindent \textbf{Time evolution for data-inferred quantile function}

\noindent In the Results section of the main text we detail how DQC is used to time evolve an analytic initial condition and how to learn the initial condition based on observations. We can also time evolve with DQC based on the observed initial quantile. The set up of the DQC is the same as when an analytic initial condition leading to Fig.~\ref{fig:analytic_res}. The result of the training is shown in Fig.~\ref{fig:obs_res}, showing that the same quality of propagation is obtained.\vspace{1mm}

\noindent \textbf{Solving reverse-time Stochastic Differential Equations with QQM}

\noindent Interesting connections between thermodynamics, machine learning and image synthesis have been uncovered in recent years. Recently, Denoising Diffusion Probabilistic Models (DDPM) \cite{SohlDickstein2015} were shown to perform high quality image synthesis at state-of-the-art levels \cite{Ho2020}, sometimes better than other generative methods like GAN-approaches. Ref. \cite{Song2020} realized that such discrete DDPMs can also be modelled as continuous processes using \textit{reverse-time} SDEs when combined with ideas from score-based generative modeling.

Ref. \cite{Anderson1982} derived the reverse-time form of general stochastic differential equations, being 
\begin{align}
\label{eq:reverseSDE}
    d \bm{X}_t = \bar{f}(\bm{X}_t, t) dt + g(\bm{X}_t, t) dW_t,
\end{align}
where now a \emph{modified} diffusion term $\bar{f}$ is given by
\begin{align}
\label{eq:reverseSDEfbar}
    \bar{f}(\bm{X}_t, t) = f(\bm{X}_t, t) - g^2(\bm{X}_t, t) \nabla_{\bm{X}} \log{ \left[ p(\bm{x},t) \right]}
\end{align}
Ref. \cite{Song2020} proposed to solve such equations with general-purpose numerical methods such as Euler-Maruyama and stochastic Runge-Kutta methods, as well as using predictor-corrector samplers. 
We envisage solving the reverse-time Fokker-Planck equation corresponding to Eq.~\eqref{eq:reverseSDE} instead, and more precisely its quantilized form, using the method described in this paper. We note that the reverse-time form (not to be confused with backward-Kolmogorov) actually looks the same, but is simply solved backwards in time starting from a `final condition' rather than an `initial condition' data set.

As noted in Ref.~\cite{Song2020}, DDPM can be regarded as the discrete form of a stochastic differential
equation (SDE). 
In Ref.~\cite{Yan2021}, a general framework based on continuous energy-based generative models for time series forecasting is established. The training process at each step is composed of a time series feature extraction module and a conditional SDE based score matching module. The prediction can be achieved by solving reverse time SDE. The method is shown to achieve state-of-the-art results for multivariate time series forecasting on real-world datasets.
These works imply that the method described here can be used for (multivariate) time-series forecasting and high-quality image synthesis.\vspace{2mm}

\noindent \textbf{Initialization of variational parameters}

\begin{figure*}
\begin{center}
\includegraphics[width=0.95\linewidth]{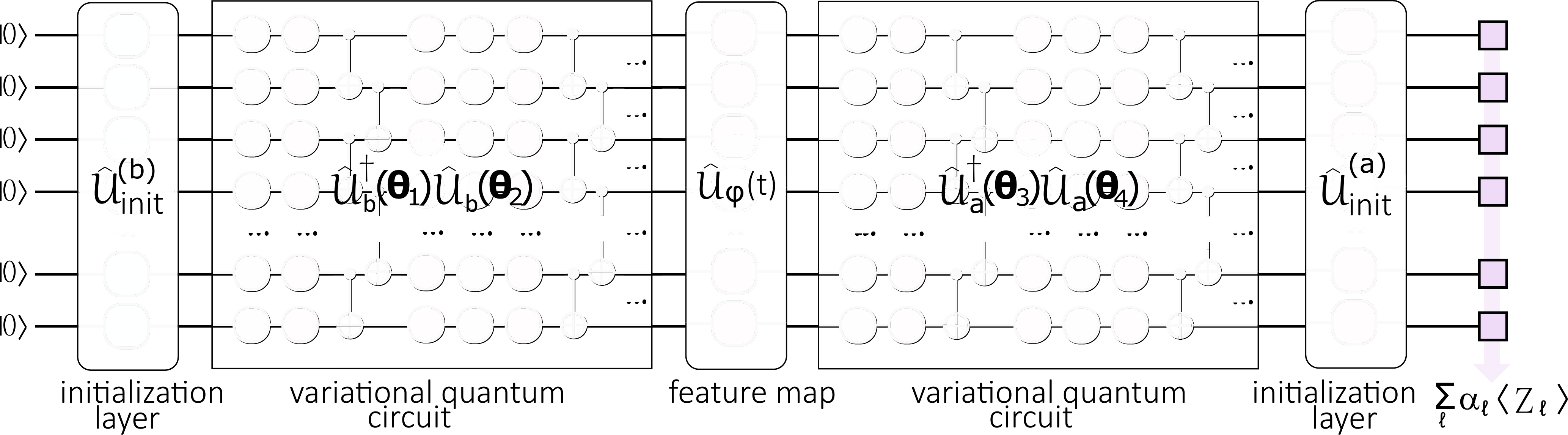}
\end{center}
\caption{\textbf{Initialization.} We show the circuit structure used for when implementing initialization. It consists of layers of rotations known as the initialization layers $\hat{\mathcal{U}}_\mathrm{init}$, variational quantum circuits made up of $\hat{\mathcal{U}}(\bm{\theta})$ and the feature map $\hat{\mathcal{U}}_\varphi (t)$. The variational quantum circuits are structured so that they can easily be initialized to the identity operator. Suitable parameters for $\hat{\mathcal{U}}_\mathrm{init}$ are chosen by performing classical function fitting. The feature map dictates which function form a basis for the fitting.}
\label{fig:init_circuit}
\end{figure*}

\noindent In this section we describe the method of \emph{parameter initialization} that we use to ensure a good starting function when implementing QCL or DQC. This is achieved by having a circuit structure where the variational circuit can be initialized to the identity operator and two layers of single-qubit rotation gates (which we refer to as the initialization layers). The parameters of the initialization layers can then be set to provide a good starting fit of the initial trial function to (an estimate of) the target function. 

The circuit structure that we use is shown in Fig.~\ref{fig:init_circuit}. The variational circuits are formed by a parameterized circuit unitary $\hat{\mathcal{U}}_{a/b}(\bm{\theta}_{k})$ followed by the circuit with the adjoint structure but independently tuned set of variational angles, $\hat{\mathcal{U}}^{\dagger}_{a/b}(\bm{\theta}_{k'})$. We include variational circuits before and after the feature map to aid expressivity. For initialization we set $\bm{\theta}_1 = \bm{\theta}_2$ and $\bm{\theta}_3 = \bm{\theta}_4$. This leads to the variational circuits being initialized as identity. We stress that during training the parameters of these circuits are considered distinct and thus when updated by classical optimiser they do not remain equal.

With the variational circuits initialized to identity the initial trial function is formed from measurements of the state received from the initialization layers and the feature map acting on the zero state. By considering a feature map which acts on each qubit individually as a product of Pauli rotations (as is the case with all feature maps discussed in this paper), and cost as the sum of individual qubit measurements, we need to treat only 1-local terms. This means that for initialization we have a circuit where the qubits are non-interacting and therefore the circuit is classically tractable. As the number of qubits $N$ increase we can still describe the system with $\mathcal{O}(N)$ at the initialization stage, and we note that in principle $k$-local terms can also be included as long as the system remains tractable. Alternatively, we can also consider the circuit based on Clifford gates. In this case the circuit remains classically tractable following the Gottesman-Knill theorem. For the remainder of this section we focus on the situation with initialization and feature map gates acting locally on single qubits. 

With the circuit structure as discussed, the general form of the initial trial function can be expressed as
\begin{align}
    f_0(t) = \sum_{j=1}^{N} \alpha_j ~ \left( c_j^{(1)}(\theta_{\mathrm{init}}) ~ g^{(1)}_j(t) + c_j^{(2)}(\theta_{\mathrm{init}}) ~ g^{(2)}_j(t) \right).
\end{align}
Here $c_j(\theta_{\mathrm{init}})$ are coefficients which depend on the angles parameterizing the initialization layer, $\alpha_j$ are the coefficients of the measurements in the cost function and $g_j(t)$ are functions of the variable encoded within the circuit by the feature map. Because this circuit is classically tractable the exact form of $c_j$ and $g_j$ remain calculable as $N$ increases. The functions $g_j$ come from the feature map encoding used. To see this note that for the feature maps considered the gate implemented on qubit $j$ is 
\begin{align}
    \hat{U}_j(t) = \hat{R}_{\beta, j}(\varphi_j(t)) = \mathrm{exp}\left( -i \hat{P}_j^\beta \varphi_j(t) / 2 \right) \\ = \mathrm{cos}(\varphi_j(t)/2) \hat{I}_j - i \mathrm{sin} (\varphi_j(t)/2) \hat{P}^\beta_j 
\end{align}
where  $\beta \in \{x, y, z\}$ are Pauli operators and $\varphi_j$ are encoding functions. Therefore, the functions $\cos(\varphi_j(t)/2)$ and $\sin (\varphi_j(t)/2)$ are introduced into the state by the feature map. The feature map encoding is thus based on the sets of functions $g^{(1)} = \{\mathrm{cos}(\varphi_j(t)) \}_j $ and $g^{(2)} = \{\mathrm{sin}(\varphi_j(t)) \}_j $. Note that the coefficients in front depend on initialization layers, the feature map and the measurement operator chosen. For instance, by choosing specific circuit structures it is possible to select only a specific set of functions to use for initialization.

Knowing the form of the initial trial function we can choose $\theta_{\mathrm{init}}$ and $\alpha_i$ such that it is a good starting state. We do this by performing classical regression. The fitting function's coefficient set $\{g_j\}$ are fitted to a limited subset of the target function values. The size of $\{g_j\}$ is at most double the number of qubits and therefore when considering NISQ implementation will be relatively small. Therefore, as a low number of fitting functions are being fitted at a limited number of points, a low-order approximation and its associated coefficients can easily be found. These coefficients are then first used to choose $\alpha_j$ such that they are of suitable scale and the coefficient magnitudes desired are reachable by $\alpha_j c_j$. With $\alpha_j$ set, the expressions $\alpha_j c_j(\theta_{\mathrm{init}})$ are then reversed to find the values which $\theta_{\mathrm{init}}$ should be initialized to.
\begin{figure}
\begin{center}
\includegraphics[width=0.95\linewidth]{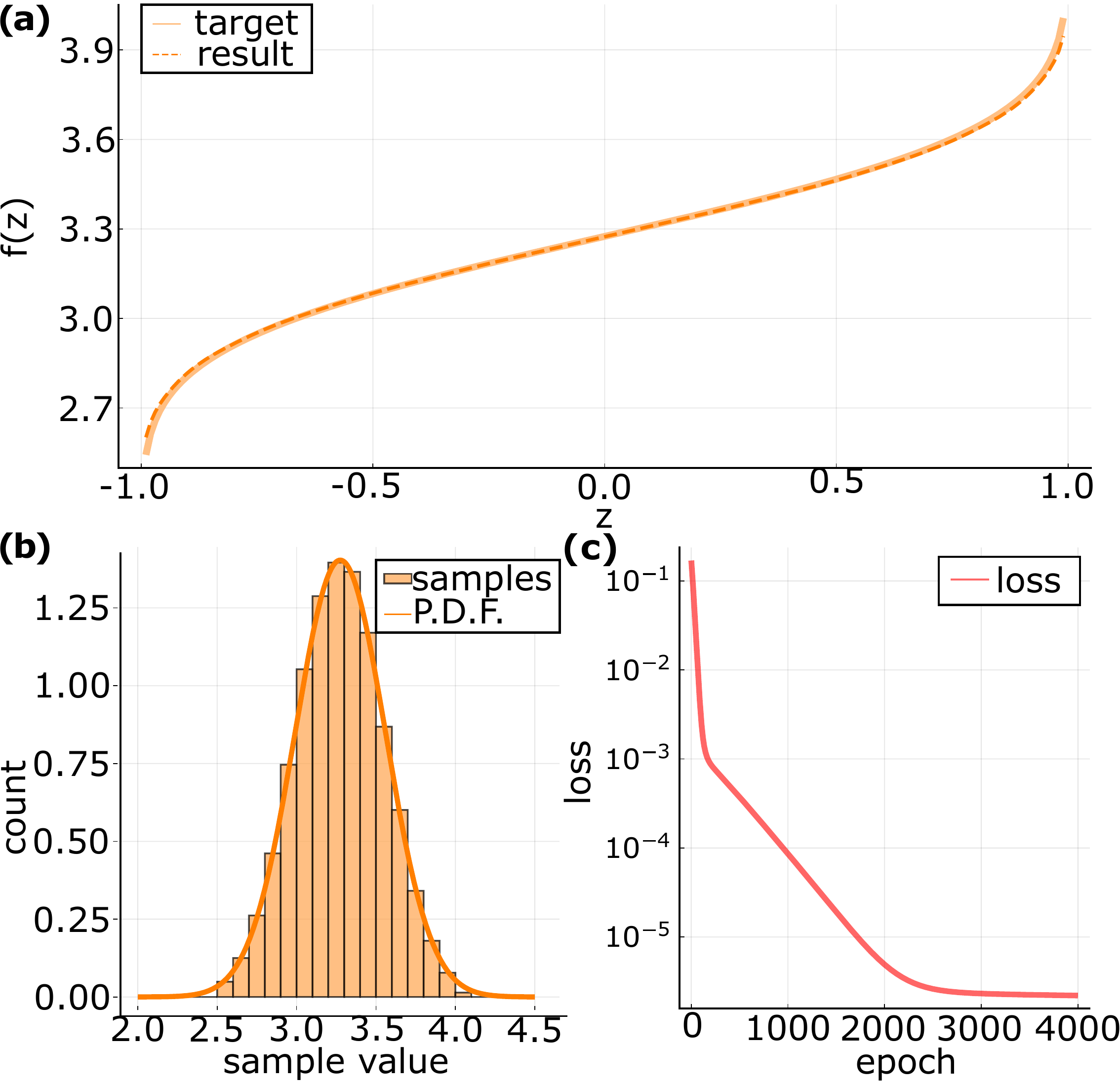}
\end{center}
\caption{\textbf{Quantile function trained on initial data --- no initialization.} \textbf{(a)} Trained QF for the Ornstein-Uhlenbeck process at $t=0$ (dashed curve labeled as \texttt{result}), plotted together with the known true quantile (solid line labeled as \texttt{target}, results overlay). \textbf{(b)} Normalized histogram of samples from the data-trained QF, plotted against the analytic distribution (PDF). $N_{\mathrm{s}}=100,000$ random samples are drawn and bin counts are normalized by $N_{\mathrm{s}}$. \textbf{(c)} Training loss at different epochs, with the final epoch producing the QF in \textbf{(a)}.}
\label{fig:noinit_qcl}
\end{figure}

Once the initialization coefficients have been calculated the standard QCL procedure is then commenced. The initialization parameters will remain fixed whilst the variational ansatz parameters will be updated. As this happens the variational ansatz will no longer remain identity operators and more fitting functions will be introduced, with the set cardinality defined by the feature map. The circuit will no longer be classically tractable. The increase in the number of fitting functions (along with number of training points) leads to a better fit of the target data being possible. When using initialization a smaller learning rate is preferred to prevent immediate divergence from the initialized function. 

We make use of initialization when training the quantile function on initial data as in Fig.~\ref{fig:quantile_qcl}. We note however that using initialization is not a requirement for convergence. In Fig.~\ref{fig:noinit_qcl} the results of training the quantile function on initial data without initialization is shown. We can see that the loss value magnitude and the fit reached is similar to that achieved when training with initialization. The difference is seen in the number of epochs --- without initialization more epochs are required to reach the same accuracy.


\begin{thebibliography}{99}
\bibitem{qqm_approach} A patent application for the method described in this manuscript has been submitted by Qu\&Co with OK, AEP and VEE as inventors.
\end{thebibliography}
\end{document}